%% file: main_version4.2-arxiv.tex
\documentclass[pdftex,twocolumn,epjc3]{svjour3} 

\usepackage{amsmath} 
\usepackage{subfig} 
\newcommand{\subfigref}[1]{(\ref{#1})}
\usepackage{siunitx} 
\usepackage{booktabs} 
\usepackage{graphicx} 
\usepackage{url} 
\usepackage{float} 
\usepackage[utf8]{inputenc} 
\usepackage{makecell}

\smartqed  

\RequirePackage{newtxtext,newtxmath} 
\RequirePackage{flushend}
\RequirePackage[numbers,sort&compress]{natbib}
\RequirePackage[colorlinks,citecolor=blue,urlcolor=blue,linkcolor=blue]{hyperref}

\journalname{Eur. Phys. J. C}

\usepackage{lineno}

\begin{document}

\title{Construction and characterisation of the DarkSide-20k veto silicon photo-multiplier tiles}

\author{The DarkSide-20k Collaboration$^{a,1}$}

\thankstext{e1}{e-mail: ds-ed@lists.infn.it}

\institute{See back for author list}

\date{Received: date / Accepted: date}

\maketitle


\begin{abstract}
Silicon photo-multipliers (SiPMs) are state-of-the-art sensors capable of detecting a single photoelectron under cryogenic conditions, with potentially lower radioactivity than widely used photomultiplier tubes. The DarkSide-20k experiment, designed to perform direct dark matter searches using liquid argon as the target material, employs SiPM technology to detect interactions in the active detector volumes, including the central dual-phase Time Projection Chamber and the Inner and Outer Veto volumes. The vetoes are designed to discriminate against radiogenic neutron and cosmic muon backgrounds associated with the dark matter search.

This paper describes the completed production and test protocols for the ``Veto Tiles'' (called \textit{vTiles}, arrays of 24 SiPMs integrated on a printed circuit board providing the power distribution and signal amplification); 16 vTiles are grouped into ``Veto Photo-Detector Units'' to instrument the Inner Veto volume.
Each vTile underwent detailed testing at room and cryogenic temperatures, confirming stable operation, high signal-to-noise ratio, and low radioactive contamination, demonstrating the robustness of the proposed design for cryogenic conditions.
The final production yield exceeded 87\%, surpassing the 80\% requirement and corresponding to 1920 Veto Tiles to populate 120 Veto Photo-Detector Units, plus an additional 6\% as spares.
\end{abstract}

\section{Introduction} \label{sec:intro} 

Although dark matter comprises 85\% of the universe's mass~\cite{Cebrian_2023,2018RPPh...81f6201R}, the nature of dark matter remains elusive. Direct detection searches aim to observe keV-scale energy deposits from dark matter scattering interactions with nuclei or electrons in sensitive terrestrial detectors~\cite{1996PhR...267..195J, LZ:2022lsv,Ramond_supersymmetry}. Non-observation to date implies dark matter particles interact rarely with baryonic matter, making the suppression of background interactions that may mimic this small signal paramount. The DarkSide-20k experiment~\cite{2018EPJP..133..131A} employs a dual-phase Time Projection Chamber (TPC) filled with 51\,tonnes of low-radioactivity liquid argon extracted from underground (UAr)~\cite{PhysRevD.93.081101} to search for dark matter interactions, with a projected sensitivity to the spin-independent dark matter-nucleon cross section down to 10$^{-48}$\,cm$^2$ at 90\% confidence level (CL) for a dark matter mass of 100 GeV/$c^2$. 

\begin{figure}[htb]
    \centering
    \subfloat[]{\includegraphics[width=0.30\textwidth]{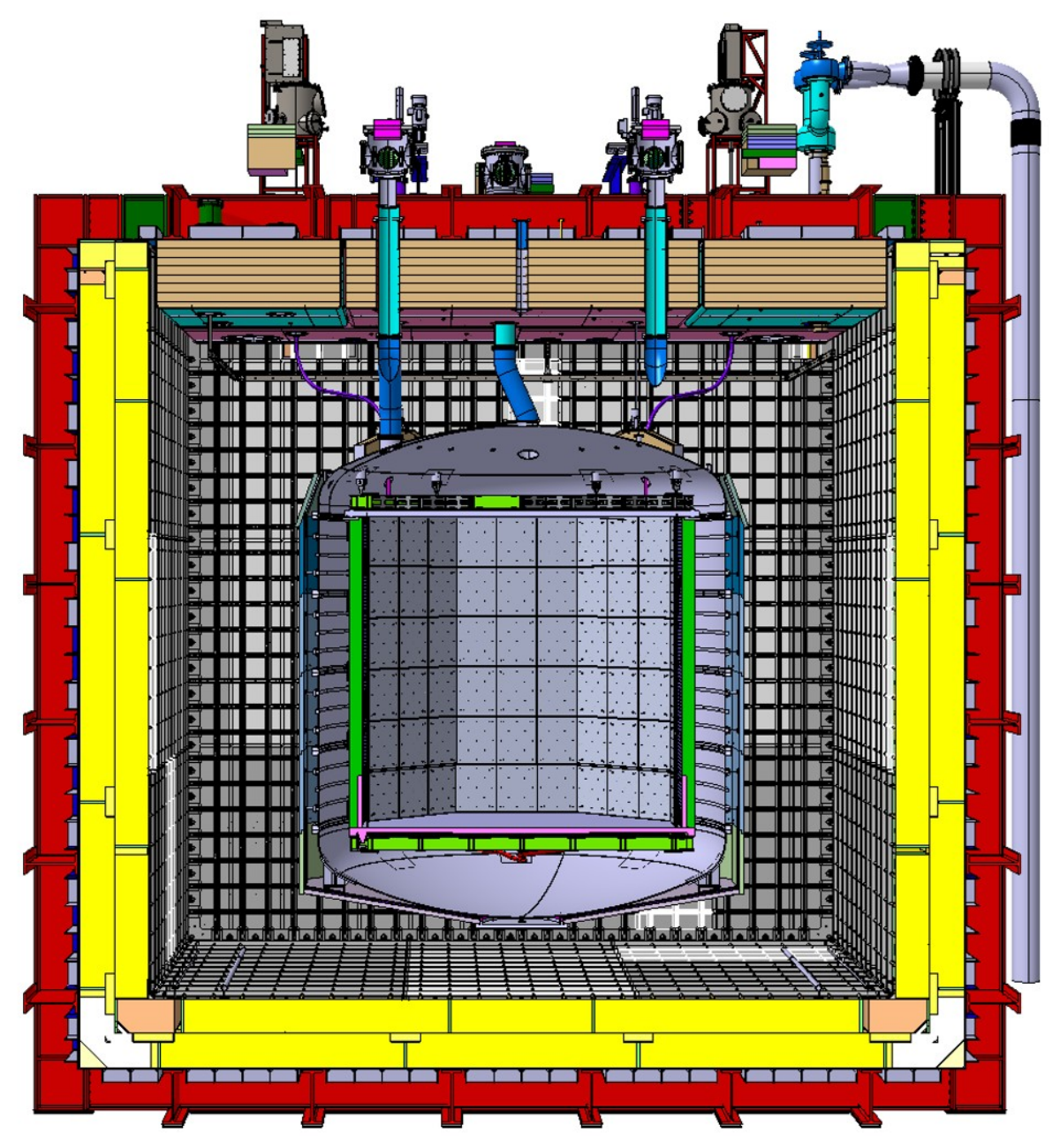}\label{fig:ds20k-a}}
    \qquad
    \subfloat[]{\includegraphics[width=0.30\textwidth]{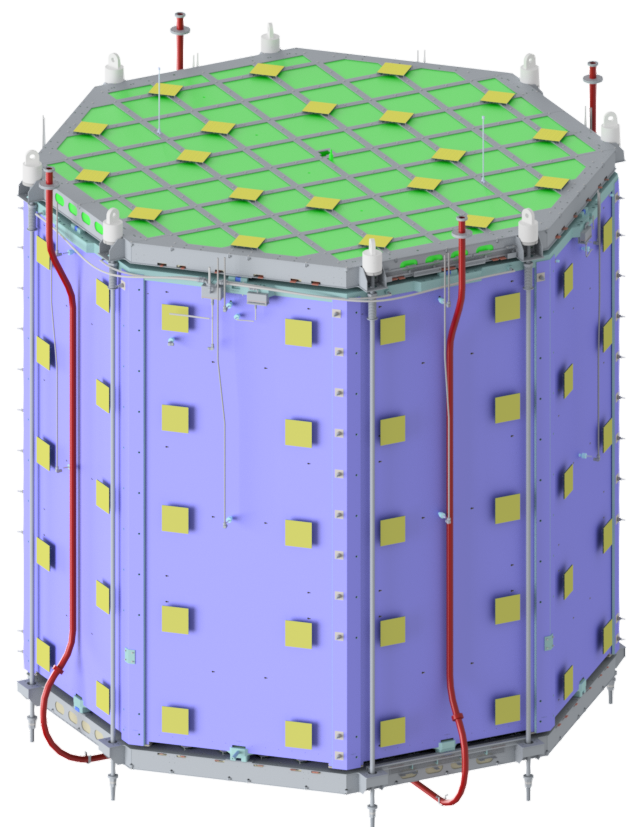}\label{fig:ds20k-b}}
    \caption{\subfigref{fig:ds20k-a} Drawings of the DarkSide-20k experiment, showing the cryostat (red and yellow structures), the Outer Veto volume and the Inner Detector inside the stainless steel vessel.
    \subfigref{fig:ds20k-b} The Inner Detector, including the upper TPC optical plane, the calibration pipes (in red) and the Inner Veto Photo-Detector Units (in yellow) attached to the PMMA walls (in blue) and at the rear of the optical plane (in green).}
    \label{fig:ds20k}
\end{figure}

A drawing of the DarkSide-20k detector is shown in Fig.~\ref{fig:ds20k}. The entire DarkSide-20k TPC is surrounded by an approximately 15\,cm polymethylmethacrylate (PMMA) shell immersed in 35\,tonnes of UAr enclosed in a stainless steel vessel (Fig.~\ref{fig:ds20k-a}); the UAr external to the PMMA shell within the vessel constitutes the Inner Veto (IV) volume.
The IV is designed to detect neutrons through their capture on hydrogen (with approximately 53\% of the total IV captures occurring in the PMMA shell) and on $^{40}$Ar.
The excited nucleus de-excites, releasing gamma rays,
depositing approximately 2.1\,MeV and 6.1\,MeV of energy respectively; gamma rays interact with the liquid argon producing secondary electrons causing scintillation; scintillation light is wavelength-shifted by polyethylene-naphthalate (PEN) films which re-emit in the optical range.
The Inner Detector (ID), formed by the TPC and the Inner Veto, is enclosed within the cryostat volume, which is filled with 650\,tonnes of liquid argon extracted from the atmosphere (AAr), comprising the Outer Veto (OV) volume.

Both the TPC and the vetoes are instrumented with large-area light detectors utilising a custom Silicon Photo-Multiplier (SiPM) technology optimised for high optical photon detection efficiency (approximately 46\% at 420\,nm, per SiPM area; paper in preparation) and low noise at liquid argon temperature~\cite{PhysRevD.93.081101, manthos2023darkside20k, Acerbi_2024}. 
The DarkSide-20k IV is equipped with 120 Veto Photo-Detector Units (called \mbox{\textit{vPDUs}}), which instrument the volume between the TPC barrel and the vessel, attached to the 8 PMMA walls and 2 caps of the barrel (yellow squares in Fig.~\ref{fig:ds20k-b}) ensuring close to full 4\,$\pi$ geometrical coverage and a uniform light response; in addition, on the PMMA walls, a 3M Enhanced Specular Reflector Film is used as a reflector and PEN acts as a wavelength shifter. The expected light yield is around 2\,photoelectrons (PEs) per keV of deposited energy.

The purpose of the IV is to reject neutron WIMP-like events by tagging neutron-induced backgrounds. Events recorded in the TPC are vetoed if a coincident signal is observed in the IV within a time window of 800\,$\mu$s. In particular, events producing more than 200\,PEs in the IV are identified as neutron-induced background.

This paper reports on the design, construction, and performance of the vTiles used to assemble the vPDUs for the DarkSide-20k experiment, to instrument the IV.

\section{Veto Photo-Detector Unit Components}\label{sec:vpdu} 

Efficient detection of scintillation light is crucial for identifying neutron captures and suppressing background events in the liquid argon. The vPDUs constitute the building blocks of the veto readout, integrating SiPMs into compact, low-radioactivity modules optimised for high photon detection efficiency and mechanical stability. 

\begin{figure*}
    \centering
    \includegraphics[width=1.0\linewidth]{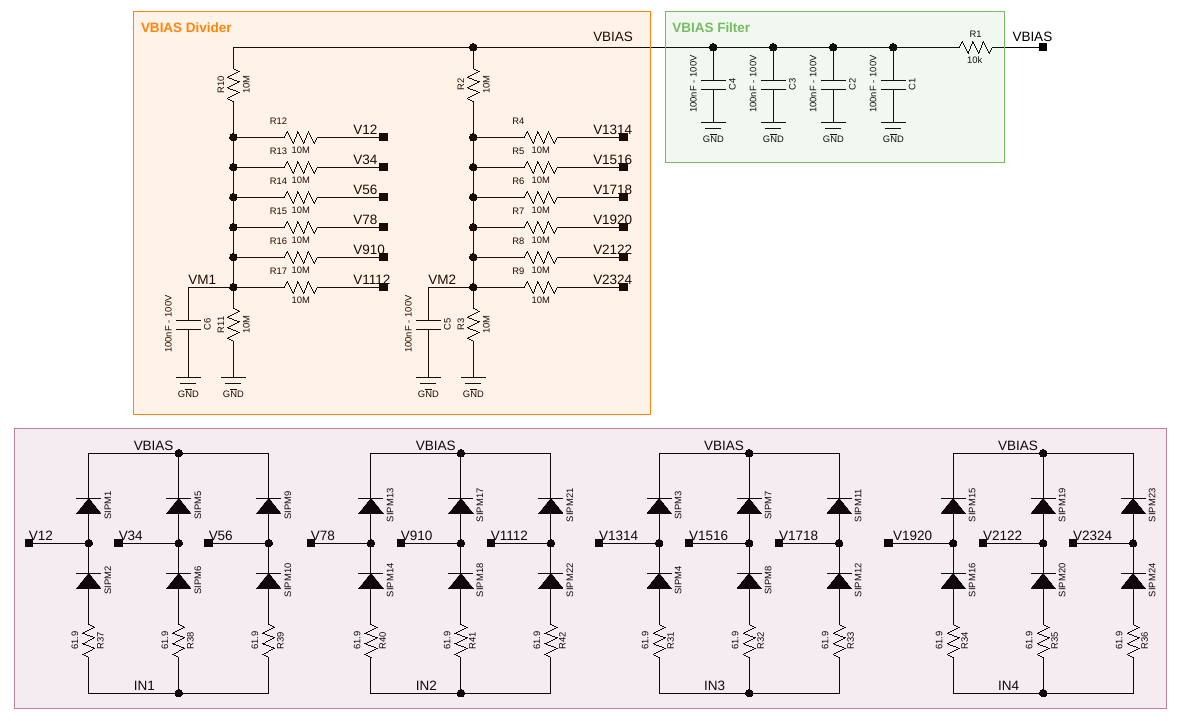}
    \caption{Schematic of the vTile electronics: it consists of 24 SiPMs arranged in a ``2s-3p'' configuration, with 3 parallel branches of 2 SiPMs in series (purple shaded region).
    A precision bias network (orange shaded region) divides the input bias voltage (VBIAS) by two to provide the mid-bias to all SiPMs; the divider is composed of two 10\,M$\Omega$ resistors (R10 and R2). The overall divider chain across all 12 bias nodes (V12 to V2324) has 10\,M$\Omega$ steps (R12 to R17 and R4 to R9) providing evenly distributed bias taps for each SiPM pair.
    Upstream of the divider, a 10\,k$\Omega$ resistor (R1) in series with VBIAS forms an RC low-pass filter (green shaded region) together with capacitors C1--C3 (100\,nF each, to ground); this filter decouples high-frequency noise from the bias supply.
    Each SiPM branch includes a 61.9\,$\Omega$ resistor (R31--R42) in series, providing frequency compensation and damping for the ASIC input. The combined output from each 2s-3p group (6 SiPMs) is routed to the corresponding ASIC channel input (IN1--IN4).}
    \label{fig:vtile-schematic}
\end{figure*}

\begin{figure}[ht!]
    \centering
     \subfloat[\label{fig:vtile-sipms}]{
        \includegraphics[width=0.42\linewidth]{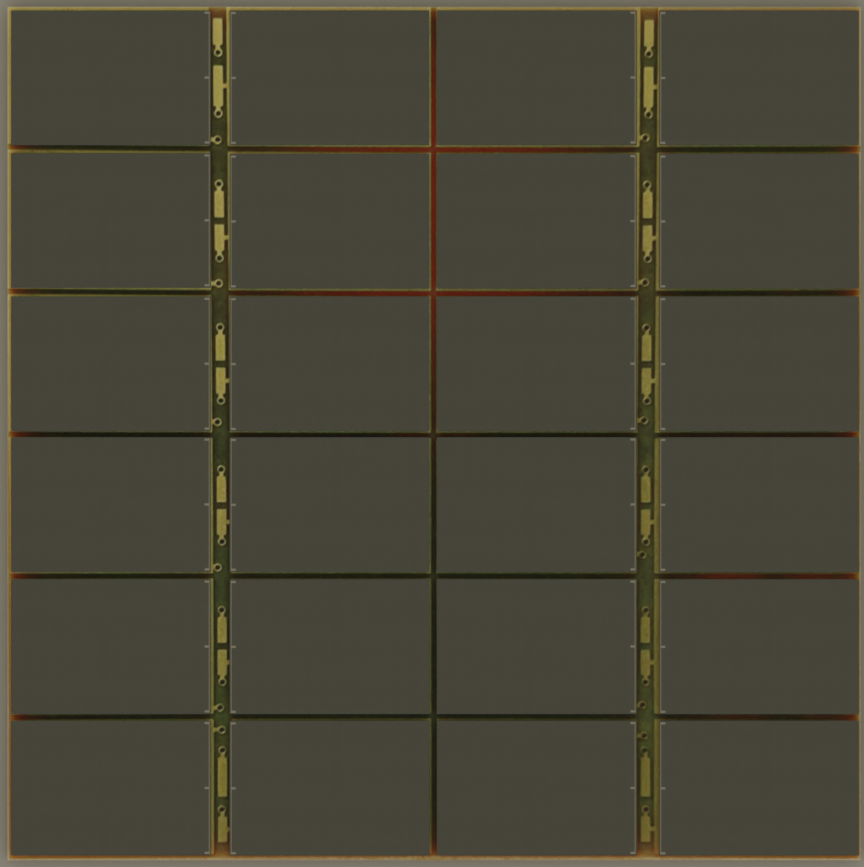}
    }
    \qquad
    \subfloat[\label{fig:vtile-circuit}]{
        \includegraphics[width=0.42\linewidth]{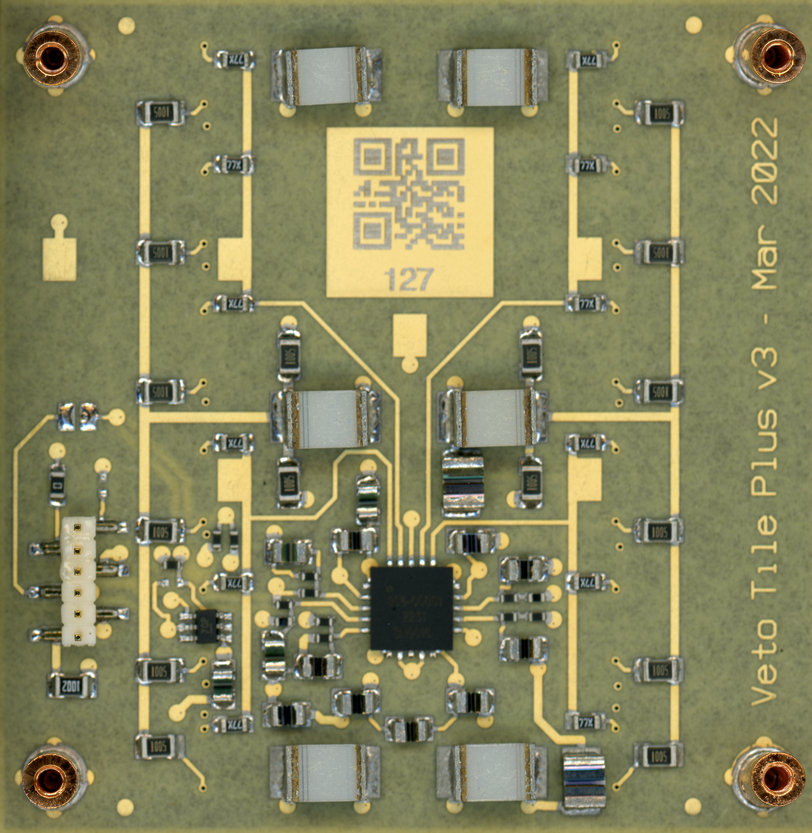}
    }
    \qquad
    \subfloat[\label{fig:Bham_tray}]{
        \includegraphics[width=0.45\linewidth]{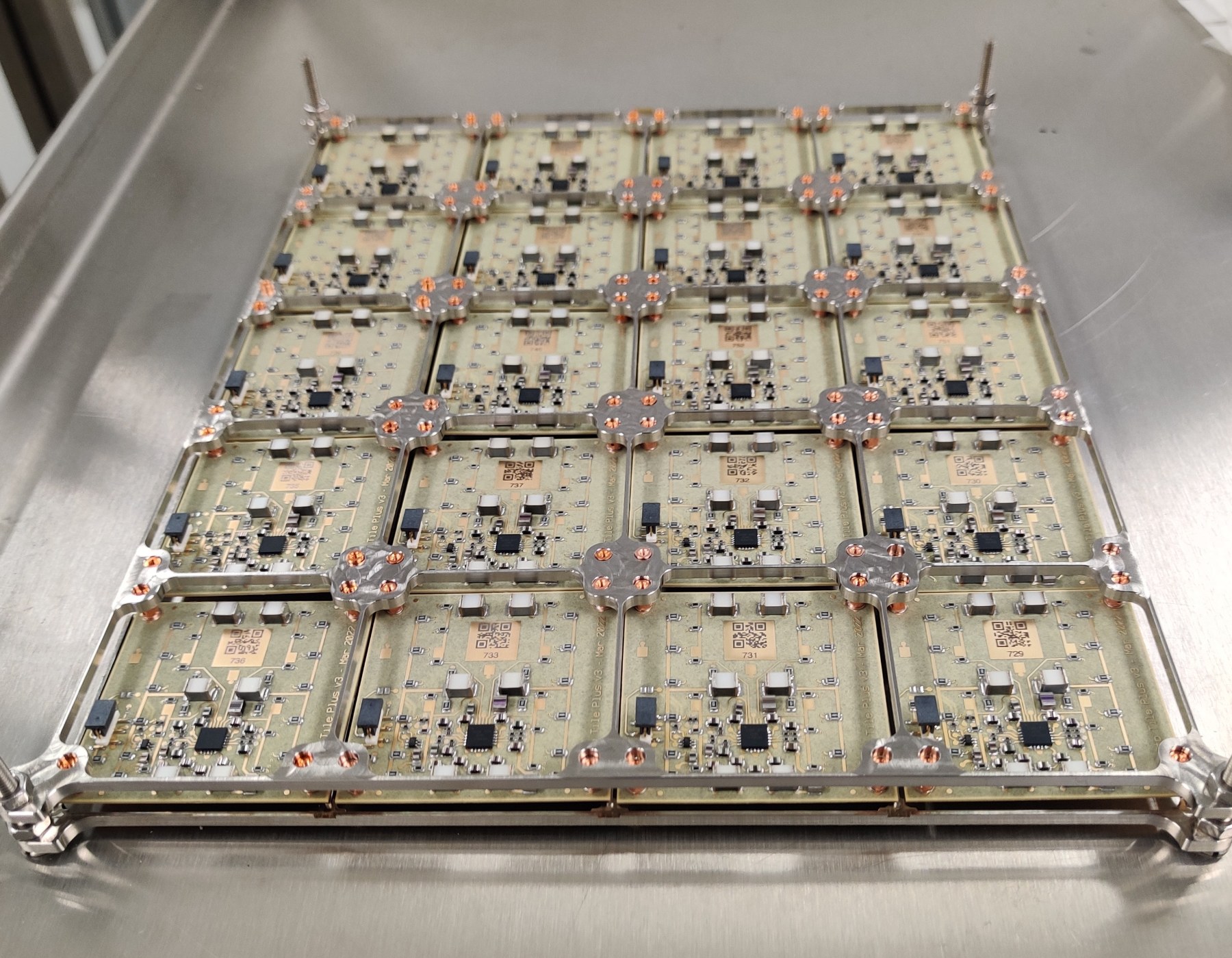}
    }
    \caption{\subfigref{fig:vtile-sipms} 24 SiPMs mounted on a vTile.
    \subfigref{fig:vtile-circuit} Populated PCB for the vTile front-end electronics. 
    The packaged ASIC amplifier is located at the bottom. The copper pillars used for the vPDU integration are present at the four corners. \subfigref{fig:Bham_tray} The holding fixture used throughout the vTile readout electronics population process.}
    \label{fig:vpcb}
\end{figure}

The fundamental unit of photon detection is an NUV-HD-Cryo SiPM, co-developed by the DarkSide-20k collaboration with Fondazione Bruno Kessler (FBK)~\cite{sipm_gola}. Each SiPM comprises 94904 Single Photon Avalanche Diodes in an 11.7$\times$7.9\,mm$^2$ area.
A total of 1400 8-inch wafers (each containing 268 SiPMs) were manufactured by LFoundry S.r.l.\ and tested by the collaboration in the ISO-6 cleanroom at the Nuova Officina Assergi facility (NOA) at the Gran Sasso National Laboratory (INFN-LNGS, Italy)~\cite{DarkSide-20k:2024usz}. Equivalent SiPM wafers are used for the TPC and veto detectors.
24 SiPMs are mounted on a 5\,cm$\times$5\,cm printed circuit board (PCB) to form an array, termed a vTile, with a single summed output signal (Fig.~\ref{fig:vpcb}). Summing of SiPM signals is done to reduce the material budget associated with cabling, without compromising the area of the readout. SiPMs in the vTiles are configured in 4 quadrants with 3 parallel branches of 2 SiPMs in series (``2s-3p'', as in Fig.~\ref{fig:vtile-schematic}), chosen to keep the capacitance at acceptable levels to preserve single photon detection performance from the relatively large area of summed SiPMs.
Population of the back side of the vTile PCB with electrical components is described in Sec.~\ref{sec:back-side-population}. Following component population, the 24 SiPMs are attached and wire-bonded to the front side of the vTile PCB, as described in Sec.~\ref{sec:front}.

\begin{figure}[ht]
    \centering
    \includegraphics[width=0.35\textwidth]{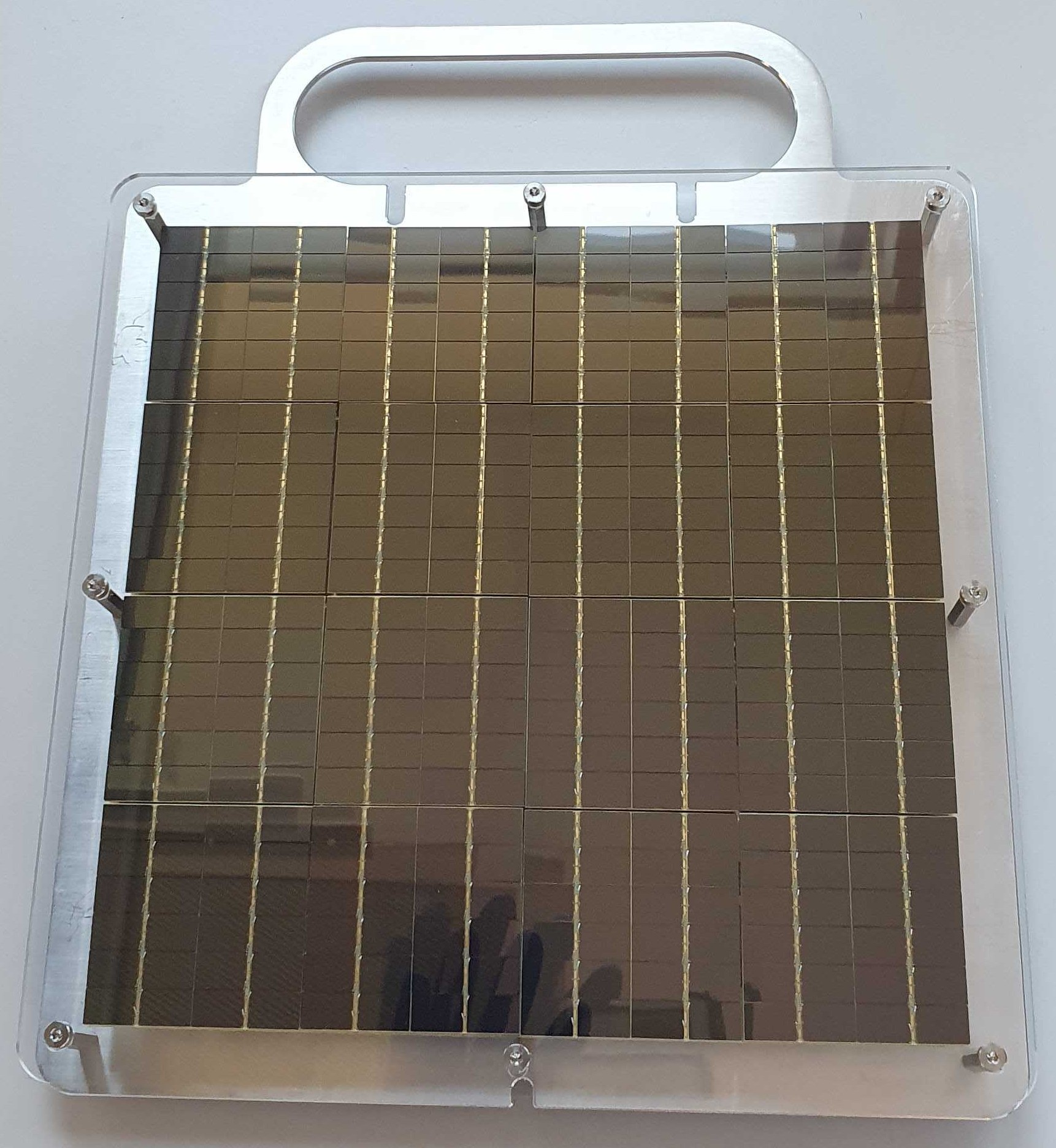}
    \caption{Fully assembled vPDU housed within the handler fixture and covered with an acrylic protection sheet.}\label{fig:vpdu}
\end{figure}

The tile substrates used in the IV are mechanically identical to those that instrument the TPC~\cite{acerbi2025productionqualityassurancequality}; however, there are differences in the electrical design as the signal amplification for the TPC tiles is performed by a single Trans-Impedance Amplifier (TIA), while for vTiles is performed by an application-specific integrated circuit (ASIC), where each quadrant has its own TIA, and the outputs from these are sent to a final stage summing amplifier which provides the vTile output~\cite{8272038}. 

16 vTiles are mounted on a 20\,cm$\times$20\,cm motherboard PCB (\textit{vMB}) to form an array with four summed output signals (Fig.~\ref{fig:vpdu}), the vPDU. Each of these output signals, termed a quadrant, sums 96 individual SiPMs covering an active area of approximately 10\,cm$\times$10\,cm. 
This summing at vPDU level reduces the number of data acquisition (DAQ) channels by a factor of 4, also limiting the radioactive budget associated with cables.
Summing of the quadrant vTile signals is implemented by the vMB, which provides the appropriate voltage bias to the vTiles, has four active adders to sum the vTile signals, and four high dynamic range differential transmitters. The vMB has 4 analogue high dynamic range differential outputs, each corresponding to the sum of 4 vTile units. 
A series of logic registers manages the onboard power system, enabling the signals and the high voltage (HV) distribution, and allowing individual vTiles and quadrants to be enabled.

\section{Production Overview}\label{sec:production}

\begin{figure*}[ht]
    \centering
    \includegraphics[clip, width=0.8\textwidth]{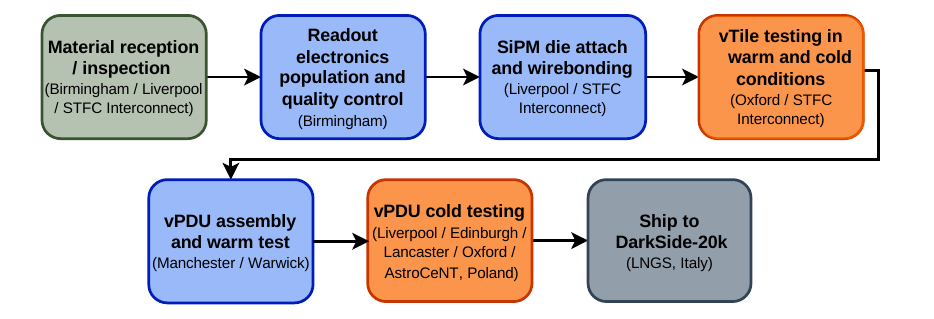}
    \caption{Production and QA/QC of vTiles and vPDUs are distributed across institutes and universities in the UK and Poland. The population of the back side of the vTile PCB with electronic components is followed by the die attach and wire bonding of 24 SiPMs on the PCB front side. The vTiles' performance is evaluated under room temperature and cryogenic conditions for compliance with the production specifications (Table~\ref{table:specs}). Subsequently, vTiles are shipped for integration, and the final step is the testing of the completed vPDUs. The qualified vPDUs are delivered to LNGS to be installed in the DarkSide-20k detector. Inspection activities are indicated in green, production in blue, testing in orange, and delivery to DarkSide-20k in grey.}
    \label{fig:flowchart}
\end{figure*}

The production, Quality Assurance and Quality Control\linebreak[4] (QA/QC) of vTiles and vPDUs are carried out by institutes and universities in the United Kingdom and Poland.
QA comprises the procedures and controls used to ensure product quality, while QC consists of the tests implemented to identify defects; QA is process-oriented and QC is product-oriented; together, they ensure the uniformity and reliability of the vTiles through early defect detection, manufacturing optimisation, and verification of operation in liquid nitrogen.
Figure~\ref{fig:flowchart} illustrates the major process steps. In brief, this process comprises engraving each PCB with a QR code for unique identification, creation of a database record to track all subsequent steps (Sec.~\ref{sec:database}), population of the back side of the vTile PCB with electronic components (Sec.~\ref{sec:back-side-population}), followed by the die attach and wire bonding of SiPMs on the PCB front side (Sec.~\ref{sec:front}). The vTile's performance is then evaluated under room temperature and cryogenic conditions (Sec.~\ref{sec:vtile_tests}) for compliance with the production specifications (Table~\ref{table:specs}). Subsequently, vTiles are shipped for integration of 16 vTiles and a vMB PCB to form a vPDU (Sec.~\ref{sec:vpdu_assembly}), and the final step is the testing of the complete vPDU under room temperature and cryogenic conditions. A future paper will describe the full characterisation and performance of the vPDUs.
Given the quantities of all the available components, the required cumulative yield from all production steps is 80\%. The completed, qualified vPDUs are shipped to LNGS to be installed in the DarkSide-20k detector's IV. Throughout the process, efforts are made to minimise the exposure of parts to atmospheric air. Parts are stored in a dry nitrogen atmosphere when not actively being processed in a cleanroom environment, and triple-bagged for all transport between cleanrooms.

\subsection{Production database}\label{sec:database}

The production database is hosted on a server at the Istituto Nazionale di Fisica Nucleare - Centro Nazionale Analisi Fotogrammi (INFN-CNAF). The database is PostgreSQL-based with replicated data and is accessible using an Application Programming Interface (API) to retrieve or insert data~\cite{Franchini:2024tut,ds20kdb}.
The production database tracks components used to build the vPDUs and records the measurements performed on the components during the various phases of the assembly in order to provide a comprehensive set of data to be used for QA/QC assessment. The database also keeps track of the location and shipment history of each component for inventory purposes.
It includes records of consumables such as ASICs and SiPM wafers, together with the test results of the devices they contain and identifiers linking each device to the components fabricated from it. Parameters of the solder materials are also tracked, including the total solder paste mass used for back-side component population and the key properties of the indium solder paste employed for SiPM attachment to the PCBs.
Assemblies of these components are catalogued alongside their electrical test results, inspection scans obtained at production stages, and recorded locations throughout the assembly workflow.

The main identifier of larger components such as vTile and vMB PCBs is a laser-engraved QR code, each of which contains numeric data fields encoded as outlined in Table~\ref{tab:qrcode}.
Some components, such as a vTile, are tracked through various stages of manufacture with a common identifier. 
QR codes were chosen for their ability to be read easily on ubiquitous consumer-grade mobile electronics. This is further supported by the free availability of open-source software libraries which facilitate automated encoding and decoding on commodity systems.
Human-readability of the encoded data is crucial to allow the swift identification of parts during the production and installation process. Numeric encoding of information stored in the QR code was chosen such that -- having read the information from the QR code using a device -- it would be easy for a person to interpret. This is augmented by the addition of a directly human-readable numeric serial number, so PCBs may be differentiated at a glance (Fig.~\ref{fig:vtile-circuit}).

For smaller components, SiPM wafers have a unique identifier engraved on them and replicated on the transport grip ring, and SiPMs are identified in the database by their original column/row location on a wafer. At the vTile level, each ASIC is identified by a unique serial number printed on its QFN-20 package. 

\begin{table}[ht!]
\caption{QR code structure, including an example for the QR code 23033003000127001 present (encoded and engraved) in
Fig.~\ref{fig:vtile-circuit}.}\label{tab:qrcode}
\centering
    \setlength{\tabcolsep}{3pt}
    \begin{tabular}{c|c|c|c|c|c|c}
    \textbf{Year} & \textbf{Month} & \textbf{Day} & \textbf{Production flag} & \textbf{Version} & \textbf{Serial} & \textbf{Part}\\
    \hline
    YY  & MM  & DD  & T/F & V.V & SSSSS & PPP \\
    23  & 03  & 30  & 0   & 3.0 & 00127 & 001 \\
    \end{tabular}
\end{table}

\subsection{Component metrology before and during vTile assembly}
\label{sec:component-metrology}

\paragraph{SiPMs} Wafers are characterised at liquid nitrogen temperature in a custom-designed probe station~\cite{DarkSide-20k:2024usz} at NOA.
The probe station measures forward- and reverse-bias current versus voltage (I-V curves) for each of the 264 accessible SiPMs on each wafer to assess breakdown voltage, quenching resistance,
leakage current before breakdown and shape of the reverse-bias I-V curve (sensitive to the correlated avalanche noise) estimated with a least-squares comparison to a reference device.

Criteria are defined to ensure a low variability of the SiPMs populating the vTiles (e.g.\ avoiding gain mismatch, pulse-to-pulse variability and ensuring similar correlated noise)~\cite{DarkSide-20k:2024usz}. In particular the breakdown voltage is required to be $27.2\pm1.0$\,V,
quenching resistance $3.35\pm1.50~\text{M}\upOmega$, 
leakage current before breakdown (at 20\,V) $\le 40~\text{pA}$ and GOF~$\le 20$ (Goodness of Fit for the reference I-V curve), with acceptance rates higher than 93\%.
SiPMs that meet the criteria are classified as ``good'' in the production database wafer map (Sec.~\ref{sec:database}) used during vTile population. Following probing, wafers are mounted onto wafer tape, diced, and shipped to the UK for production of vTiles.

\begin{figure*}[ht!]
\centering
\subfloat[\centering pad flaw affecting wire bonding]{\includegraphics[width=0.25\textwidth]{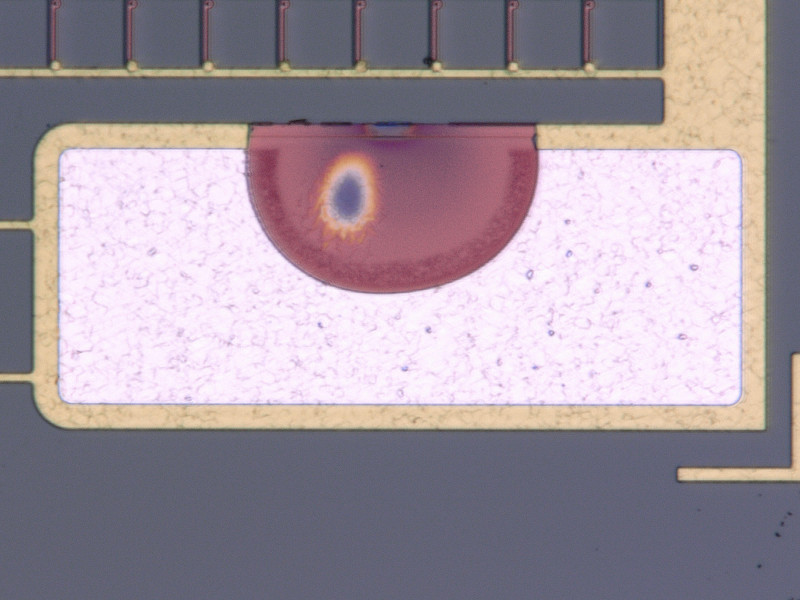}}
\label{fig:subfig1}
\qquad
\subfloat[\centering pad flaw affecting connectivity]{\includegraphics[width=0.25\textwidth]{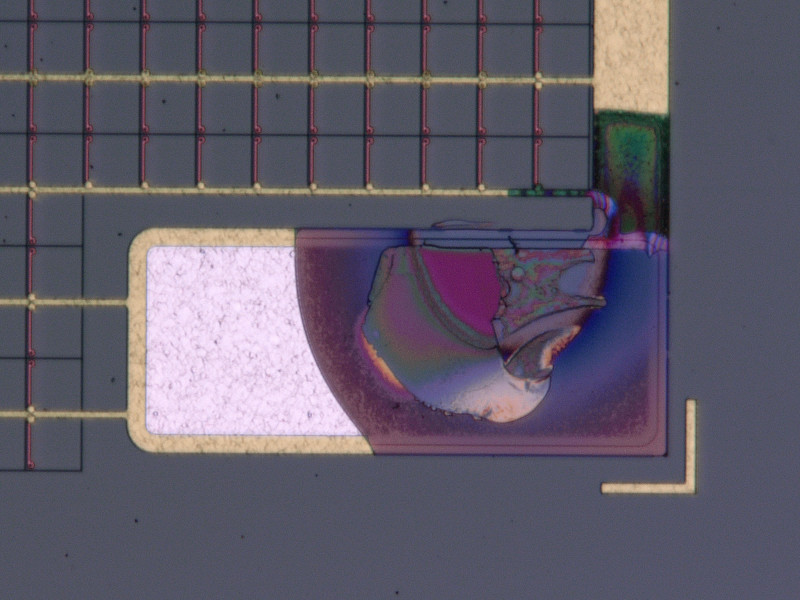}}
\label{fig:subfig2}
\qquad
\subfloat[\centering break in metal between pads]{\includegraphics[width=0.25\textwidth]{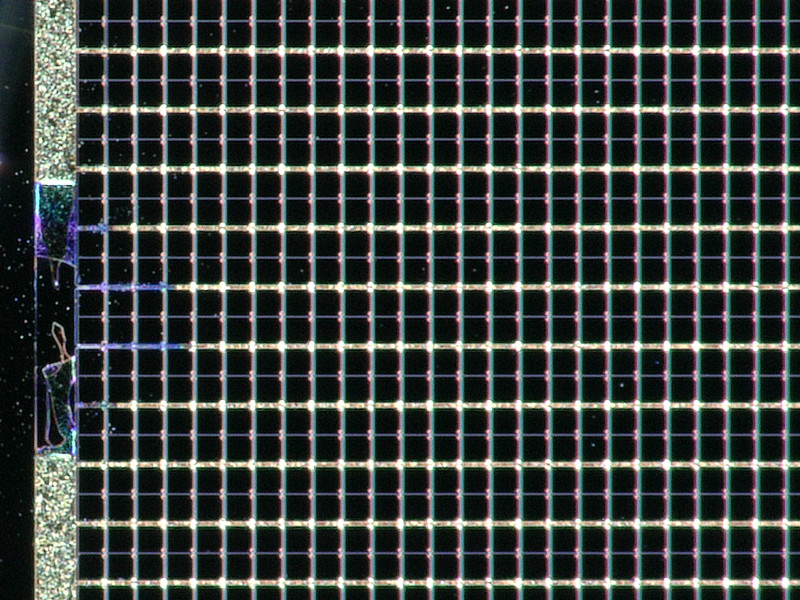}}
\label{fig:subfig3}

\subfloat[\centering surface deposit]{\includegraphics[width=0.25\textwidth]{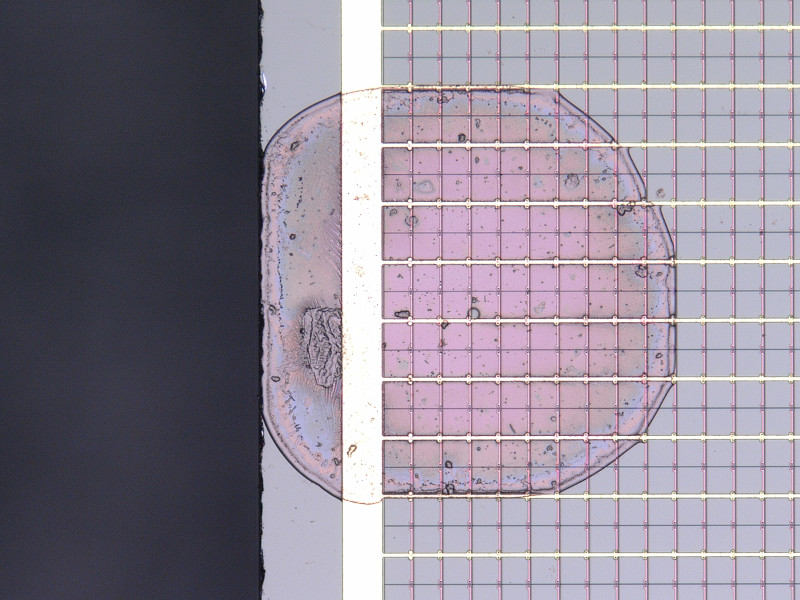}}
\label{fig:subfig4}
\qquad
\subfloat[\centering metal discolouration]{\includegraphics[width=0.25\textwidth]{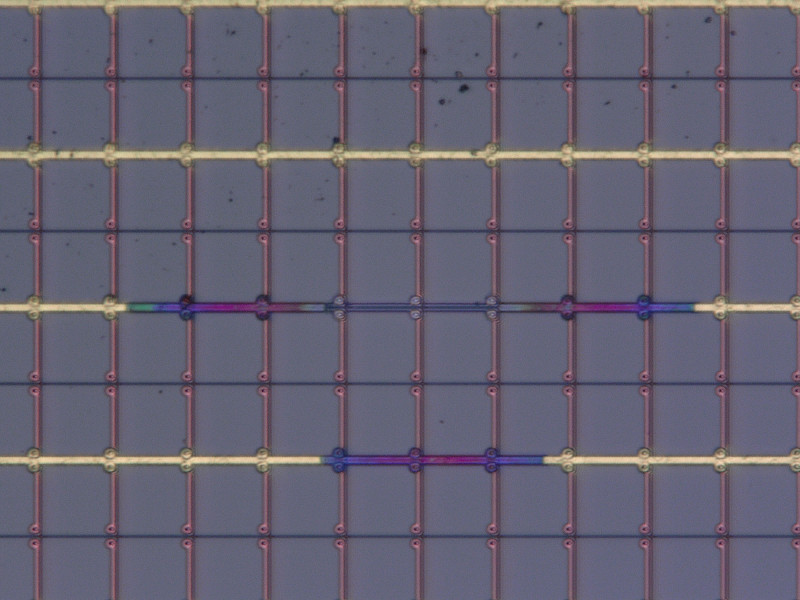}}
\label{fig:subfig5}
\qquad
\subfloat[\centering minor metal break]{\includegraphics[width=0.25\textwidth]{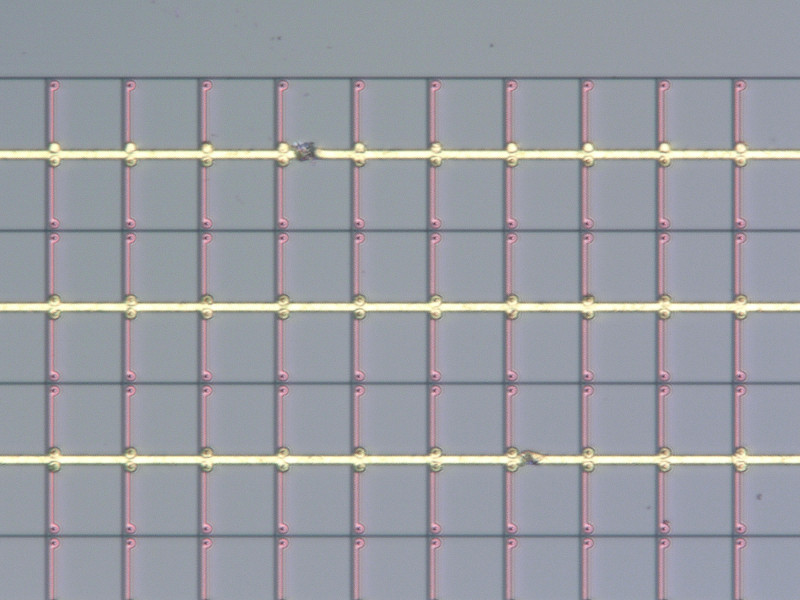}}
\label{fig:subfig6}
\caption{Surface features from SiPM visual inspection. The top row shows classes of defects that lead to SiPM rejection, as they are likely to cause problems during integration. In contrast, the bottom row displays minor defects that show no significant correlation with known pathologies during testing.}
\label{fig:sipm_inspection}
\end{figure*}

After cryogenic testing at NOA, wafers are diced and dispatched to assembly sites (Sec.~\ref{sec:vpdu}), where a visual inspection of SiPMs is carried out at the stage of wafer picking. 
SiPMs may have notable physical flaws but still pass tests at NOA, not showing up in the predominantly electrical tests. Such gross flaws may cause problems with wire-bonding, or disconnect a region of SPADs, which has shown a correlation to I-V double-breakdowns.
The top row of Fig.~\ref{fig:sipm_inspection} shows classes of flaws that would cause SiPMs to be rejected at this stage. On the left is an example of a small pad flaw that would create problems for the connection of the SiPMs to the PCB's pads (wire bonding), given that the entire pad area is used to accommodate the source heels of two wire bonds. 
The centre image shows a more extensive flaw that prevents connectivity to the bulk of the device, even if a wire bond was placed on the viable pad area. The right image shows a break in the track connecting the three SiPM pads, limiting the connection to a portion of the device. 

The number of SiPMs deemed unsuitable for production per wafer varies. From 238 production wafers (92\% of the total Veto production), 17.6\% of SiPMs were deemed unsuitable
based on the manufacturer’s wafer map (electrical test, visual inspection), downstream wafer testing before dicing (electrical test) and visual inspection.

Production wafers have shown themselves to be free of gross defects such as large-scale physical damage, scratches, and discolouration. Minor observable surface features such as those shown in the bottom row of Fig.~\ref{fig:sipm_inspection} feature on approximately 20\% of SiPMs that had passed their electrical tests.
These mild flaws have shown no significant correlation to known pathologies in testing, nor any degradation of the photon detection efficiency. Therefore, we judge these to be acceptable for production, but we still log their presence for record-keeping purposes and later analysis.

\paragraph{vTile components} The QA/QC protocol for the vTile PCB prior to component population includes visual inspection, metrology of the copper pillars that are used for mechanical integration of the vTiles with the vMBs, and functional test of the ASIC. 

Visual inspection is done to ensure the edge quality of the PCB meets the IPC-A-600 standard (specifying the absence of residual fibres at the edges), to reject PCBs showing pad discolouration, or PCBs with $x$-$y$ dimensions out of specification (49.5$\pm$0.1\,mm for each side). 

The performance of 3188 ASICs was evaluated, at room temperature, by charge injection emulating a single PE signal input to ensure functionality before an ASIC is mounted to a vTile PCB, as reported in Ref.~\cite{ROGERS2024169723}.
The selection criteria for acceptable ASIC response correspond to the design current draw of 32\,mA and are defined from the distribution of rise time versus signal amplitude; devices were required to fall within the 95\% confidence contour (elliptical region) of this distribution, ensuring consistent timing and gain characteristics across all ASICs.

\begin{table*}[ht!]
\small
\centering    
\caption{DarkSide-20k specifications and Quality Assurance and Quality Control criteria for accepting components to instrument the IV, showing the yields of the different stages of the production.
The vTile CR tests and the vMB tests did not assess the physics performance of the detector, as they were purely functional electronic checks to confirm that the boards were operating correctly.}
\begin{tabular}{l|cc|c}
    \hline
    \textbf{Quantity} & \textbf{Min} & \textbf{Max} & \textbf{Yield} \\
    \hline
    \multicolumn{3}{l}{\textbf{vTile ASIC tests} (Sec.~\ref{sec:component-metrology})}  & 95\%\\
    \hline
    ASIC current draw (@2.5\,V) [mA] & 30 & 34 & 
    \\
    \makecell[l]{ASIC Charge Injection\\(rise time vs.\ peak amplitude confidence ellipse)} & \multicolumn{2}{c|}{95\%} & \\
    \hline
    \multicolumn{3}{l}{\textbf{vTile CI tests} (Sec.~\ref{sec:back-side-population})}  & 98\%\\
    \hline
    vTile Charge Injection & & \\ (99\% of the peak amplitude distribution) [mV] & 837 &  913  \\
    \hline\hline
    \multicolumn{3}{l}{\textbf{vTile CR tests} (Sec.~\ref{sec:cr})} & 99\% \\
    \hline
    Capacitance/quadrant (6 SiPMs) [nF] & 31.07 & 32.93 \\ 
    Resistance/quadrant (6 SiPMs) [$\Omega$] &  26.8 & 28.6   \\
    \hline
    \multicolumn{3}{l}{\textbf{vTile warm tests} (Sec.~\ref{sec:breakdown-measurement})} & 97\% \\
    \hline
    RMS [mV] (@20\,V) & 1.0 & 1.4 \\
    Breakdown [V] & 66 & 68 \\
    \hline
    \multicolumn{3}{l}{\textbf{vTile cold tests @69\,V} (Sec.~\ref{sec:vtile_tests})} & 91\% \\ 
    \hline
    1-PE amplitude [mV]     &  3.4 & 4.8  \\
    RMS [$\mu$V]            &  300  & 450  \\
    SNR                     &  8 & -- \\
    PCR [Hz/mm$^2$]         &  0    & 1.8  \\
    \hline\hline
    \multicolumn{3}{l}{\textbf{vMB tests} (Sec.~\ref{sec:vpdu_assembly})} & 98\% \\ 
    \hline
    Current drawn per quadrant (at warm) [mA] &  46 & 48\\
    Total current drawn (at warm) [mA]        &  163 & 167 \\  
    Current drawn per quadrant (at cold) [mA] &  25 & 26\\
    Total current drawn (at cold) [mA]        &  105 & 106 \\
    LV current drawn per vTile (@7\,V, at warm) [mA] &  71 & 73\\ 
    HV current drawn per vTile (@40\,V, at warm) [$\mu$A] &  3.894 & 3.898\\ 
    \hline
\end{tabular}
\label{table:specs}
\end{table*}

\paragraph{Automated imaging} Image capture is performed at the entrance and exit of production stages in order to provide condition records that can be compared before and after all transport steps, and to measure particulate density. Image capture hardware was selected based on an assessment of image quality of a scan of a reference USAF1951 target object. The comparison between an Epson v850 scanner and a Canon 90D DSLR camera was favourable to the scanner in this context. On the scanner hardware, little additional visual quality was observed beyond 3200\,DPI. At this resolution, wire bonds are clearly resolved and file size with lossless compression for a 5\,cm square image is tolerable at approximately 60\,MiB per scan.

\begin{figure*}[ht!]
  \centering
    \includegraphics[width=0.7\linewidth]{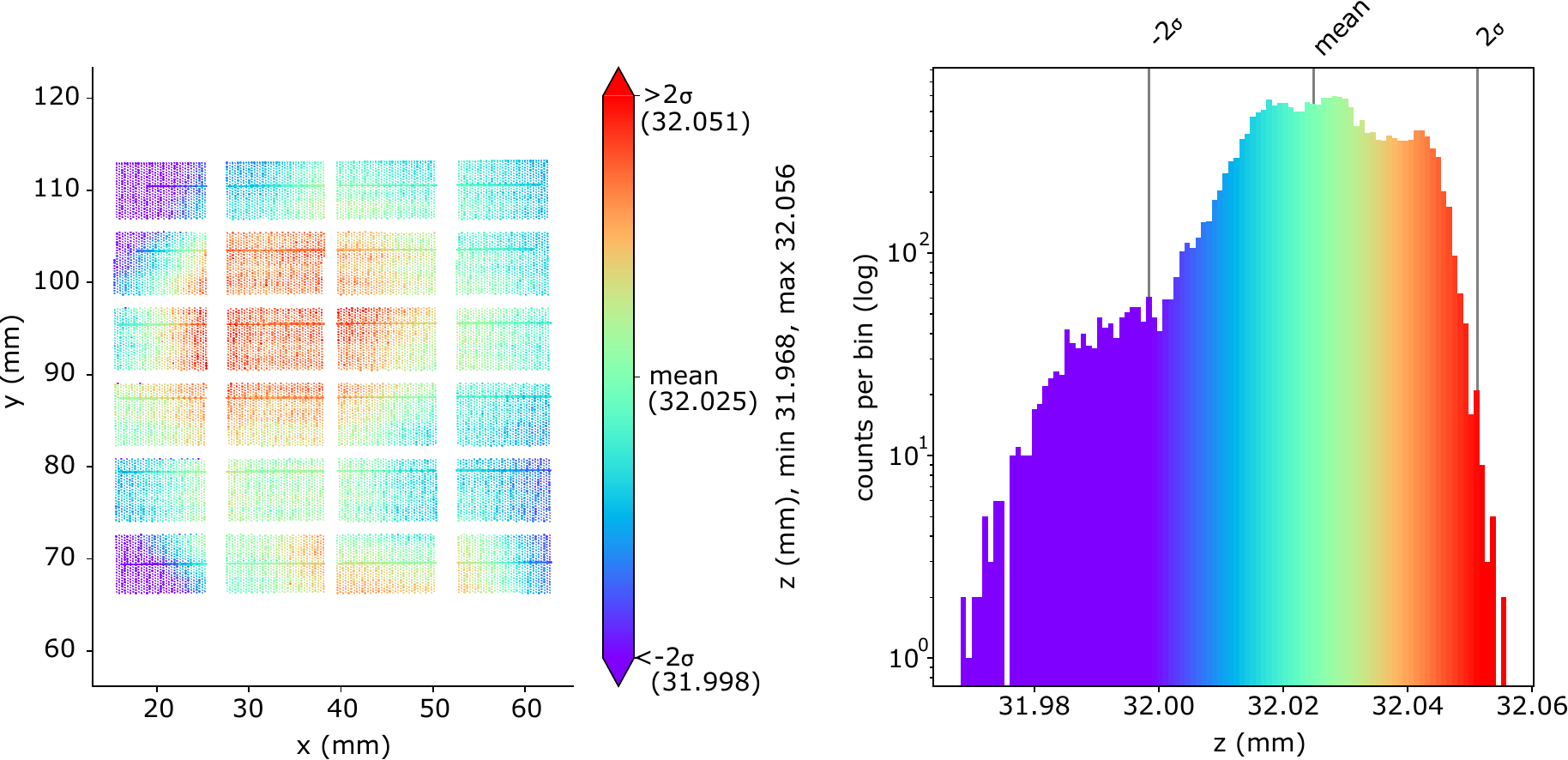}
    \caption[Flatness plot for a pre-production DarkSide-20k vTile]{Flatness plot for a pre-production DarkSide-20k vTile (QR code 22061703000024001), showing a worst-case SiPM surface $z$-axis deviation of 72\,$\mu$m.}
    \label{fig:flatness}
\end{figure*}

Higher resolution automated imaging is performed periodically, on a subset of assembled vTiles, using an OGP Inc.\ SmartScope 624: an example scan is shown in Fig.~\ref{fig:flatness}. This is done to ensure consistent assembly, verifying the uniformity tolerance of the front SiPM surface, which has been generally~$\le$100\,$\mu$m. Deviations may arise from PCB curvature and/or SiPMs not being level in comparison to the PCB substrate to which they are bonded.

\subsection{vTile PCB component population}
\label{sec:back-side-population}

The workflow for the population of the vTile PCB with the front-end electronics components starts by registering the unique QR code of each PCB in the database. PCBs passing acceptance metrology are then individually loaded onto a manual stencil printer (ESSEMTEC SP-002) using a 0.10\,mm thick stencil, and 0.15$\pm$0.1\,g of Type-4, lead-free, solder paste (CHIPQUIK SMDLTLFP250T4, melting point 138$^\circ$C), compatible with the radiopurity requirements (Table~\ref{tab:components}), is applied to the PCB footprint.

Subsequently, PCBs are loaded into an automatic pick-and-place (Mechatronika M60). The use of a milled stainless-steel tray enables population of up to 20 PCBs concurrently. All components used to populate the vTile PCB are stored in ISO-6 cleanroom environment in sealed anti-static bags. After checking the positioning of the PCB fiducial marks with the software checklist function, placement of the 67 components begins. This procedure requires approximately 70\,minutes for 20 PCBs. PCBs are then visually inspected for missed/misplaced components. If any are identified, the PCB is placed back into the pick-and-place and the specific component is repositioned on the PCB. 

To ensure the proper placement of the vTile PCBs on the vPDU PCB, within a 100\,$\mu$m tolerance, copper pillars are soldered at the 4 corners of the vTile (see Fig.~\ref{fig:vtile-circuit}). 
To meet the alignment specification and to avoid pillars moving out of alignment during the reflow process, a custom stainless-steel jig is fixed on top of the tray into which PCBs are placed, shown in Fig.~\ref{fig:Bham_tray}. After pillar placement, the populated PCBs with the pillar alignment jig are reflowed.

\begin{figure}[ht!]
    \centering
    \subfloat[\label{fig:temp_prof}]{
        \includegraphics[width=0.7\linewidth]{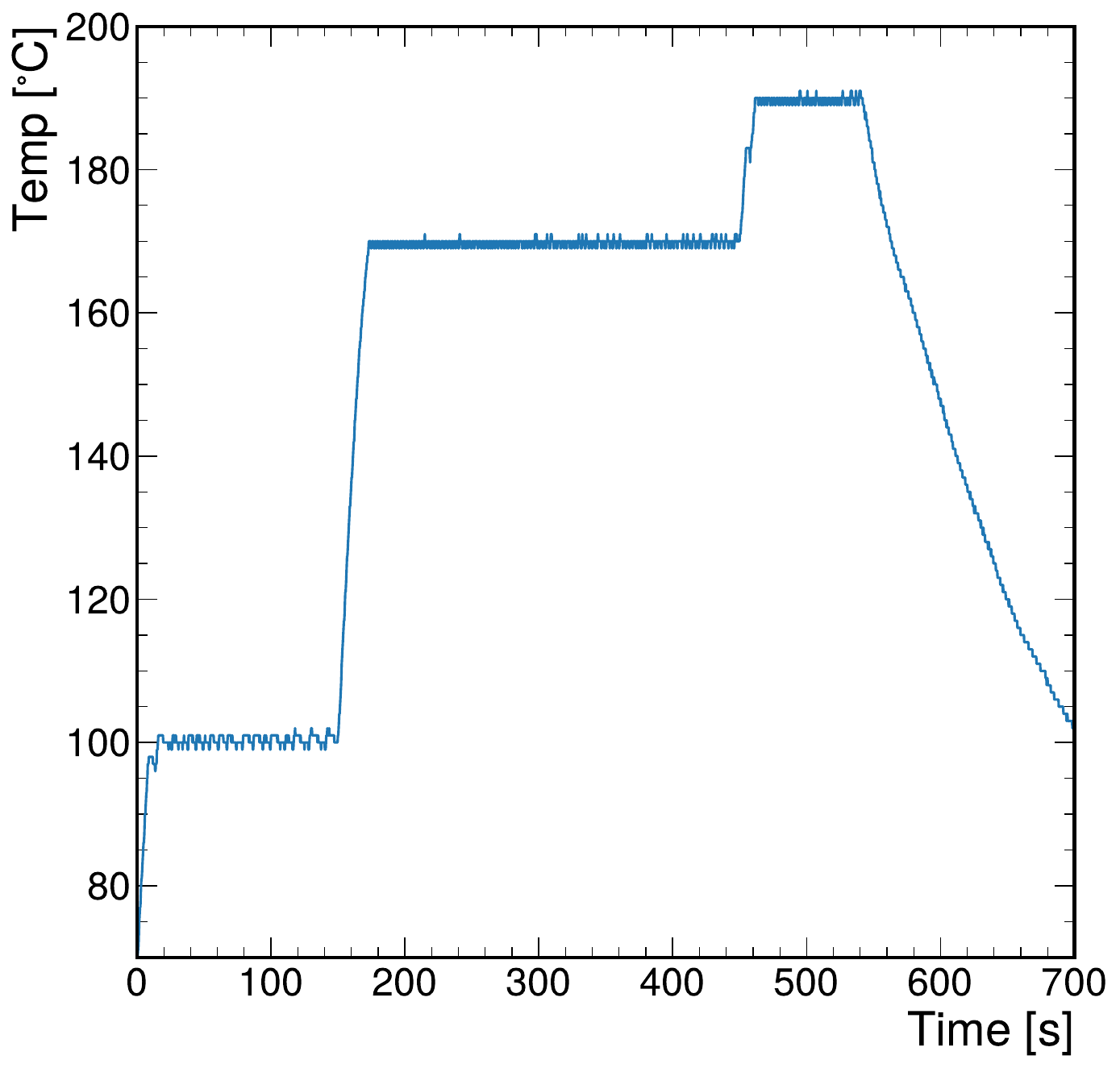}
    }
    \qquad
    \subfloat[\label{fig:vTile_amp}]{
        \includegraphics[width=0.7\linewidth]{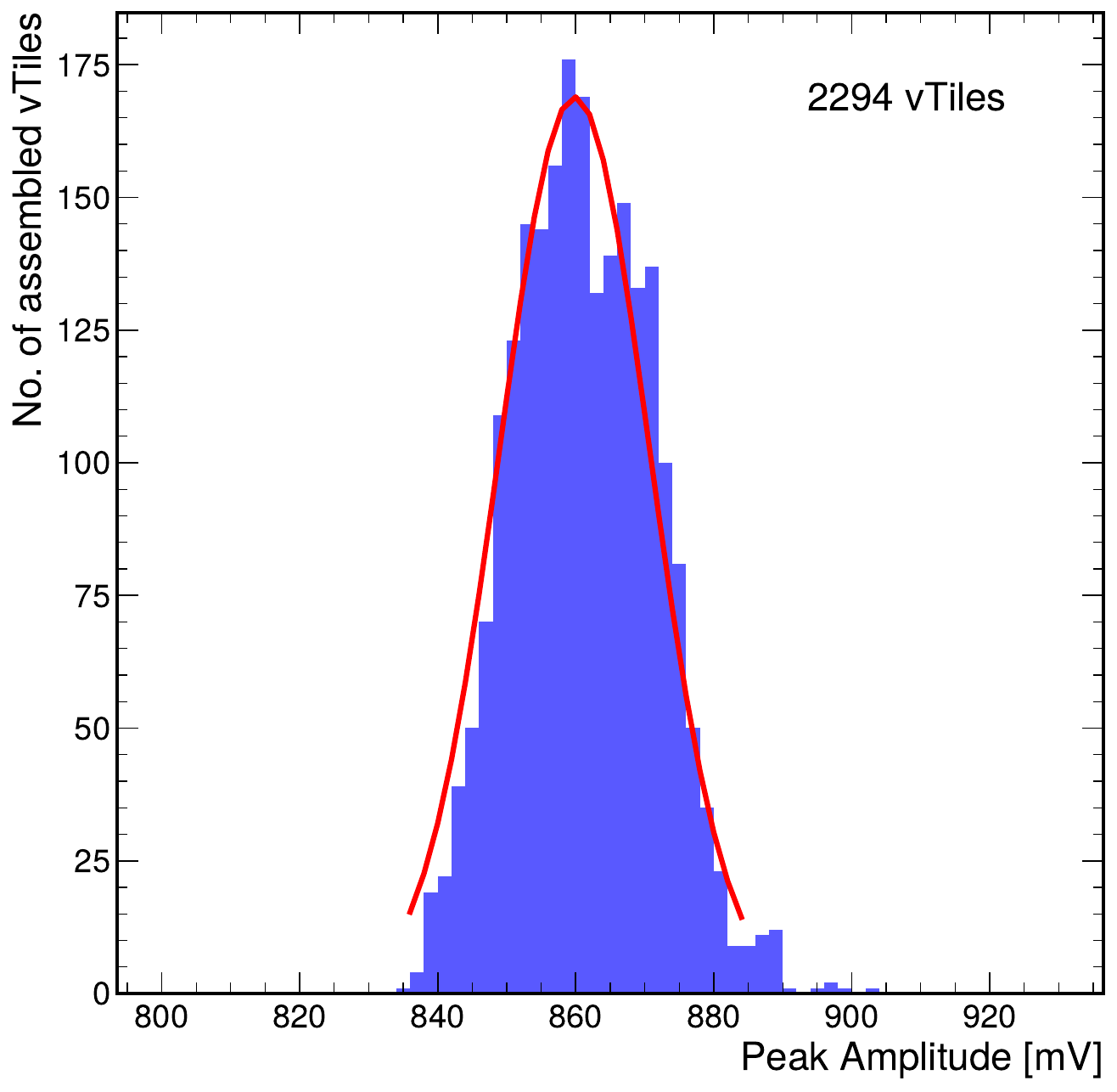}} 
    \caption{\subfigref{fig:temp_prof} Temperature profile used in the forced convection oven. \subfigref{fig:vTile_amp} Peak amplitude distribution of the response to a square test pulse for vTiles populated with readout electronics, fitted with a Gaussian.}
    \label{fig:vtile-bham}
\end{figure}

Reflow is done in a forced convection oven (C.I.F. FT05 Advanced). The reflow temperature profile includes 150\,s at 100$^\circ$C, 300\,s at 170$^\circ$C, and 90\,s at 190$^\circ$C, as shown in Fig.~\ref{fig:temp_prof}. This temperature profile is shifted to higher temperatures compared to the spec sheet value for the solder, in order to evaporate possible contaminants in the solder paste. The temperature is monitored by 3 thermocouples throughout reflow, and recorded for future reference. Finally, a high-resolution image of the populated side is taken with a scanner (described in Sec.~\ref{sec:component-metrology}) to verify the vTile condition and enable traceability for potential damage identification.

Electrical acceptance tests are done by injecting a test pulse into the PCB (100\,mV, 4\,$\mu\textrm{s}$, 1\,kHz square pulse), and recording the output pulse amplitude of the populated vTile. 
The response of the assembled vTiles followed a Gaussian distribution, as shown in Fig.~\ref{fig:vTile_amp}. To ensure uniform performance across the photodetectors, 1.8\% of vTiles exhibiting anomalous responses (very low gain, or showing no gain) -- thus not contributing to the Gaussian distribution in Fig.~\ref{fig:vTile_amp} -- were excluded from subsequent analysis. Since no silicon was consumed for these rejected units, they were not included in the production yield calculation. The results for all tested vTiles were recorded in the database.

\subsection{vTile PCB front side population}
\label{sec:front}

The operations that populate the front side of the vTile, die attach and wire-bonding, are performed at two sites (at the University of Liverpool and STFC Interconnect) using the same tooling, to provide a degree of robustness in the production logistics in case of downtime at one site. 
Parts storage and assembly operations are done in an ISO-5 cleanroom environment, whilst packaging in vacuum bags for dispatch is done in an ISO-7 area of the cleanroom. 

SiPMs are picked from the diced wafers using the wafer map of acceptable devices resulting from cryoprobing, and passing visual inspection as described in Sec.~\ref{sec:component-metrology}. This is performed promptly using a die ejector (Semiconductor Equipment Corporation model 4800), to minimise the time the SiPMs spend attached to wafer tape. Picked SiPMs are transferred to trays of 24 SiPMs that mimic the physical arrangement of SiPMs on the vTile to ease later assembly stages. 
During the picking process, a database record is created that documents the relationship between SiPM position in the tray and the original wafer column/row location, to provide traceability information for each die. 

To verify PCB condition before assembly commences, scans are taken of the populated back side, and the unpopulated front side, using the automated imaging protocol described in Sec.~\ref{sec:component-metrology}.
Prior to die attach, a database entry is created for the vTile to verify the fitness of solder and SiPMs for production use before they are permanently bonded to the vTile. The database submission software warns if any SiPMs have been previously allocated to another vTile, and whether all SiPMs are of production standard. In addition, it checks that the solder has not passed its expiry date and has not been at room temperature for more than 30 days.

\paragraph{Die-attach} The PCB is inserted into a vacuum jig, and indium solder paste (Indium Paste NC-SMQ80 Ind\#1E\linebreak[4] 52In48Sn Type 4 83\%) is applied using a laser cut stencil, achieving a 19$\pm$1\,mg wet mass. The indium solder paste has a maximum temperature reached on the temperature profile lower than the solder paste used for the PCB components. The laser-cut stencil allows the deposition of 3 discrete solder dots for each SiPM, evenly distributed along the horizontal centre line of each SiPM. This arrangement minimises the probability of gas pockets forming within the indium solder volume, which may apply stress to the joint during the transition from cryogenic temperatures. SiPMs are then placed using an alignment stencil.
At this point, any removable surface particulates are lifted away from the SiPM with fine sticky swabs designed for optics, to avoid attachment to the surface during baking.

To achieve the required goal of having wires that separate SiPMs of less than 300\,$\mu\textrm{m}$ thickness, stencil apertures for SiPMs are created using a chemical etching process. SiPMs are bonded to the front side using a conductive indium solder that has a lower reflow temperature (118$^\circ$C) than the solder used to attach the back-side components (approximately 138$^\circ$C). This enables the PCB to endure a second re-flow cycle to bond the SiPMs without releasing components from the back side. Baking is performed at a temperature of 132$^\circ$C in an ARGO LAB TCN30 Plus convection oven. Temperature logs are collected from 3 thermocouples, 2 in the oven and 1 measuring ambient temperature; these logs are stored for future traceability.

Shear tests have been performed on 3 pre-production vTiles for a total of 36 SiPMs. A checker board pattern of 12 SiPMs was placed on each PCB, to allow space for the shear testing head, and SiPM movement during the test, without interfering with other dies. After glue deposition and SiPM placement, the vTiles were reflowed with the standard procedure described above.
Once the vTiles had cooled, they were placed in the chuck of the XYZtec Sigma shear tester, and the 100\,kgf shear head was selected. The head was manually moved into position to one of the short edges of each SiPM, and the machine performed a touchdown, raised the shear head 50\,$\mu \textrm{m}$ above the touchdown point, and then moved a total of 2000\,$\mu$m along a single axis, perpendicular to the short edge of the SiPM.
The results gave both average and peak force, but the main point of interest was the peak force; almost no elastic deformation was observed: the stress builds quickly to peak force and then an almost immediate transition to plastic deformation, the yield point, leading sometimes to an explosive disconnection of the SiPM from the PCB.
The results were satisfying, with the majority of SiPMs shearing around 10--12\,kg of force, and the lowest results still in excess of 8\,kg, more than adequate for the vTile production.

\paragraph{Wire bonding}

Once the SiPMs have been attached, the vTile proceeds to wire bonding. This is a largely manual process performed on a HESSE BJ820 wire bonder using Al-1\%Si, 25\,$\mu\textrm{m}$ thick wire. Bond parameters used are Ultrasonic 18\%, bond force 22\,cN, loop height from source 350\,$\mu\textrm{m}$, apex (loop shape) 80\%, stop after deformation 34\% (provides foot width of approximately 8.5\,$\mu\textrm{m}$). For redundancy, 2 bond wires per SiPM pad are used. Despite the 550\,$\mu\textrm{m}$ height difference between the SiPM and PCB pads, wire bonding is possible without the need for a deep access head. 

The in-plane distance between SiPM and PCB pads is variable due to the placement tolerance for SiPMs. A distance of not less than 200\,$\mu\textrm{m}$ between the conductive edge of the SiPM and the closest of the two bond wires is maintained. This allowance reduces the probability of unintended contact between bond wires and the SiPM edge, should their relative position change during the transition to cryogenic temperatures.

\begin{figure}[ht!]
    \centering
    \includegraphics[width=0.95\linewidth]{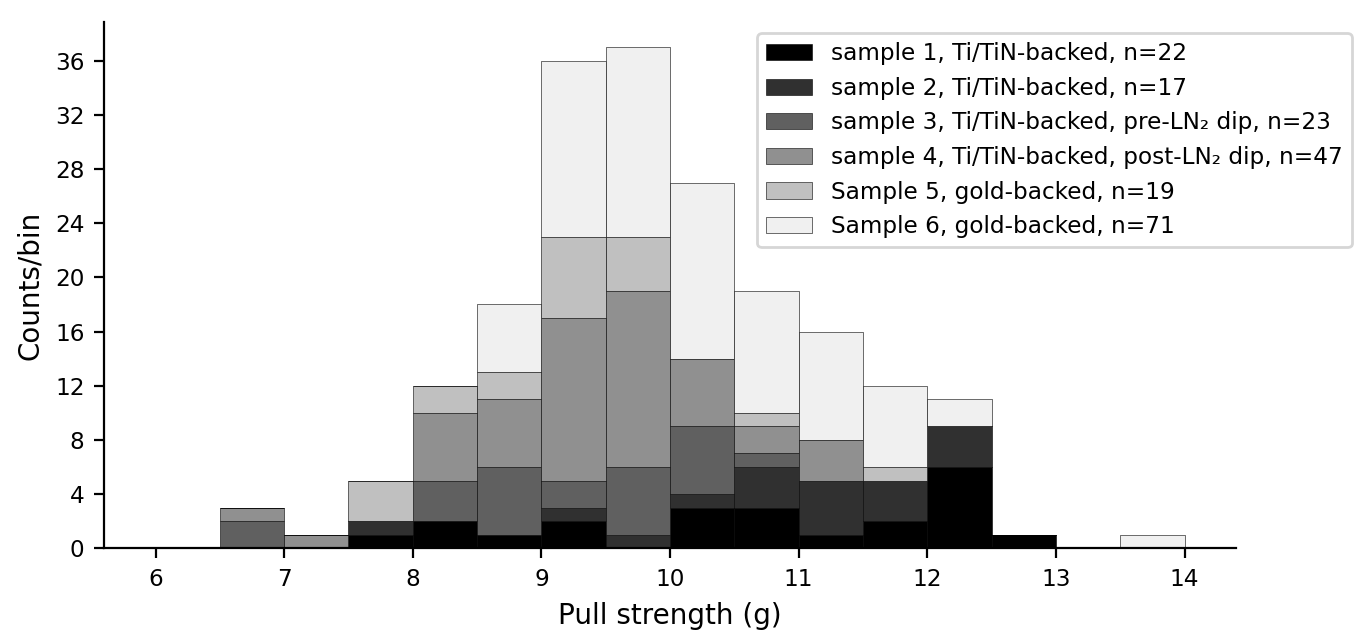}
    \caption{Wire bond pull strength test.}
    \label{fig:pull_test_str}
\end{figure}

During the pre-production R\&D stage, 
pull tests of bond strength vs.\ length were performed on selected test pieces.
Measured wire bond pull strengths for test pieces are summarised in Fig.~\ref{fig:pull_test_str}.
Wire bond breaks were satisfactory, with the bond wire either remaining intact and separating cleanly from the pad at either end, or breaking over the span. No instances of pad separation -- from either PCB or SiPM -- were observed (Table~\ref{table:pull_test_break_types}). Minimum break strengths were deemed acceptable, as these exceed the minimum acceptable threshold of 5\,grams (Table~\ref{table:pull_test_break_types_stat}).
%
\begin{table}[ht!]
    \footnotesize
    \renewcommand{\arraystretch}{1.4} 
    \centering
    \begin{tabular}{c|c|c|c|c|c|c}
         \toprule 
         \textbf{Break Type} & \multicolumn{6}{c}{\textbf{Sample}} \\\cline{2-7}
         & \textbf{1} & \textbf{2} & \textbf{3} & \textbf{4} & \textbf{5} & \textbf{6} \\
         \midrule 
         No Recording$^A$ & 0 & 0 & 0 & 0 & 0 & 2 \\
         Source heel break$^B$ & 14 & 8 & 19 & 37 & 13 & 0 \\
         Dest heel break$^C$ & 8 & 9 & 4 & 10 & 6 & 67 \\
         Source foot lift$^D$ & 0 & 0 & 0 & 0 & 0 & 0 \\
         Dest foot lift$^E$ & 0 & 0 & 0 & 0 & 0 & 0 \\
         Span break$^F$ & 0 & 0 & 0 & 0 & 0 & 2 \\
         \bottomrule 
    \end{tabular}
    \caption{Wire bond pull tests: break types. $^A$ user error; $^B$ bond separated at the SiPM pad; $^C$ bond separated at the PCB pad; $^D$ bond peeled from SiPM pad; $^E$ bond peeled from PCB pad; $^F$ wire bond broke.}
    \label{table:pull_test_break_types}
\end{table}

\begin{table}[ht]
    \footnotesize
    \renewcommand{\arraystretch}{1.4} 
    \centering
    \setlength{\tabcolsep}{1pt}
    \begin{tabular}{c|c|c|c|c|c|c}
         \toprule 
         \textbf{Statistic} & \multicolumn{6}{c}{\textbf{Sample}} \\\cline{2-7}
         & \textbf{1} & \textbf{2} & \textbf{3} & \textbf{4} & \textbf{5} & \textbf{6} \\
         \midrule 
         min value [g] & 7.926 & 7.702 & 6.835 & 6.891 & 7.581 & 8.609 \\
         max value [g] & 12.683 & 12.365 & 10.835 & 11.218 & 11.748 & 13.797 \\
         SD [g] & 1.52 & 1.21 & 1.09 & 0.89 & 1.06 & 1.02 \\
         SD [\% of mean] & 14.3 & 11.04 & 11.82 & 9.42 & 11.45 & 9.95 \\
         \bottomrule 
    \end{tabular}
    \caption{Wire bond pull tests: statistics of break types.}
    \label{table:pull_test_break_types_stat}
\end{table}

\paragraph{Post-die attach testing}\label{sec:cr}
vTiles progress to capacitance/resis\-tance (CR) testing after the 24 SiPMs have been attached and wire-bonded.
The test system makes use of a Keysight E4980A Precision LCR meter to measure the capacitance and resistance of each of the 4 quadrants of a vTile (constituted by 6 SiPMs). 
vTiles are mounted to a custom test PCB with a 3D printed socket inside a dark enclosure. Measured CR values are checked against reference figures encoded within the test software. 
The CR test quickly identifies problems with electrical connections (wire bonds, indium solder joints); the acceptance criteria are reported in Table~\ref{table:specs}.
CR testing enables efficient production by weeding out problematic devices as early as possible in the production process.

Finally, any visible surface particulates are removed from the SiPM surfaces, and a scan is taken of the front surface with SiPMs and wire bonds visible.

At this point the vTiles are packaged for shipping (to cryogenic vTile testing), and the remaining database entry tasks are completed: the vTile and its SiPMs, then CR-test results are recorded; subsequently, scanned images are losslessly compressed and uploaded to a shared filestore and matching database entries are constructed.

\subsection{Transportation and storage fixtures}
The materials and structures used for shipping serve three purposes: to prevent surface contamination of the electronics destined for the experiment due to radon progeny, to minimise the probability of physical damage, and to act as a reliable storage medium while assets await processing or installation.

\begin{figure}[ht!]
    \centering
    \includegraphics[width=0.9\linewidth]{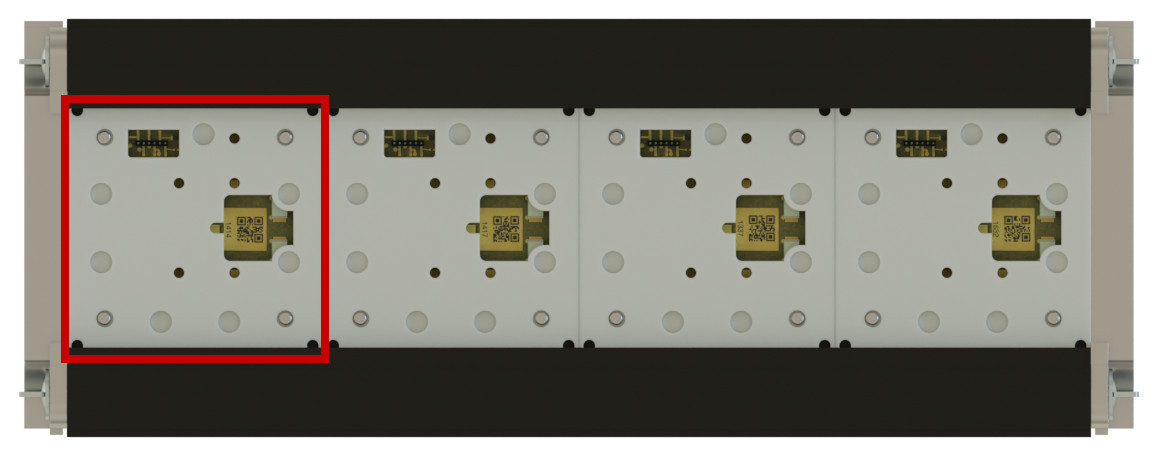}
    \caption{vTile shipping fixture (underside) containing four vTiles mounted in their process carriers; identifying QR codes remain visible. The red outline indicates a single 6\,cm$ \times $6\,cm process carrier.}
    \label{fig:vsfu}
\end{figure}

For small electronic assemblies the structural protection in direct contact with electronics, is provided as follows.
During the vTile assembly 2-piece trays, which are 3D printed and cleaned in an ultrasonic bath before use, are used to ship up to 12 vTiles and ensure safe transportation, as shown in Fig.~\ref{fig:Bham_tray}. This prevents any damage to the back-side components, including the connector pins and threaded mounting pillars, which are most prone to damage due to their height above the PCB surface.
Fully assembled vTiles are shipped in a stackable carrier structure that protects the wire bonds and the upper surface of the SiPM, and limits the potential for particles to settle on the upper surface. The QR codes for each vTile remain visible while in the carrier. Each fixture contains four vTiles mounted in their process carriers. These are secured in place using end pieces attached with quick-release Dzus fasteners (Fig.~\ref{fig:vsfu}).
The final assembled vPDUs are housed in a structure formed from a stainless steel handler plate, standoffs, and a clear acrylic sheet to protect the top side of the 16 vTiles with their exposed wire bonds (Fig.~\ref{fig:vpdu}).

Each of the three packaging assemblies described above is then enclosed in a series of three partial-vacuumed bags. The innermost bag is an ESD moisture barrier bag (type 1), while the outer two are polyethylene/polyamide bags assessed for their impermeability to radon (type 2).
The full process consists of sealing the item in a type 1 bag;
sealing the type 1 bag, including a desiccant sachet and a moisture monitor strip, inside a vacuumed type 2 bag; finally adding a second vacuumed type 2 bag with a desiccant sachet and a moisture monitor strip inside.
The triple-bagging process provides a degree of redundancy against bag and/or heat seal joint failure. The moisture monitor strip provides notice of outer bag failure, while the presence of desiccant provides short-duration limited environmental control should a seal fail.

\subsection{vTile performance characterisation}\label{sec:vtile_tests}

Each vTile is characterised at room (``warm'') and at cryogenic (``cold'') temperatures in order to establish whether it has single photoelectron (SPE) detection and noise performance sufficient to proceed to integration into a vPDU, and, to ensure the uniformity of the production process over time. 
Uniformity and stable performance across all vTiles are critical for DarkSide-20k, as the veto system relies on consistent photon detection efficiency and low noise over hundreds of channels to achieve reliable background rejection. Variations in gain or noise among vTiles could introduce systematic differences in light yield across the detector, degrading the overall veto efficiency and, ultimately, dark matter sensitivity.

\begin{figure}[ht!]
    \centering
    \subfloat[Testing mount]{\includegraphics[width=0.30\textwidth]{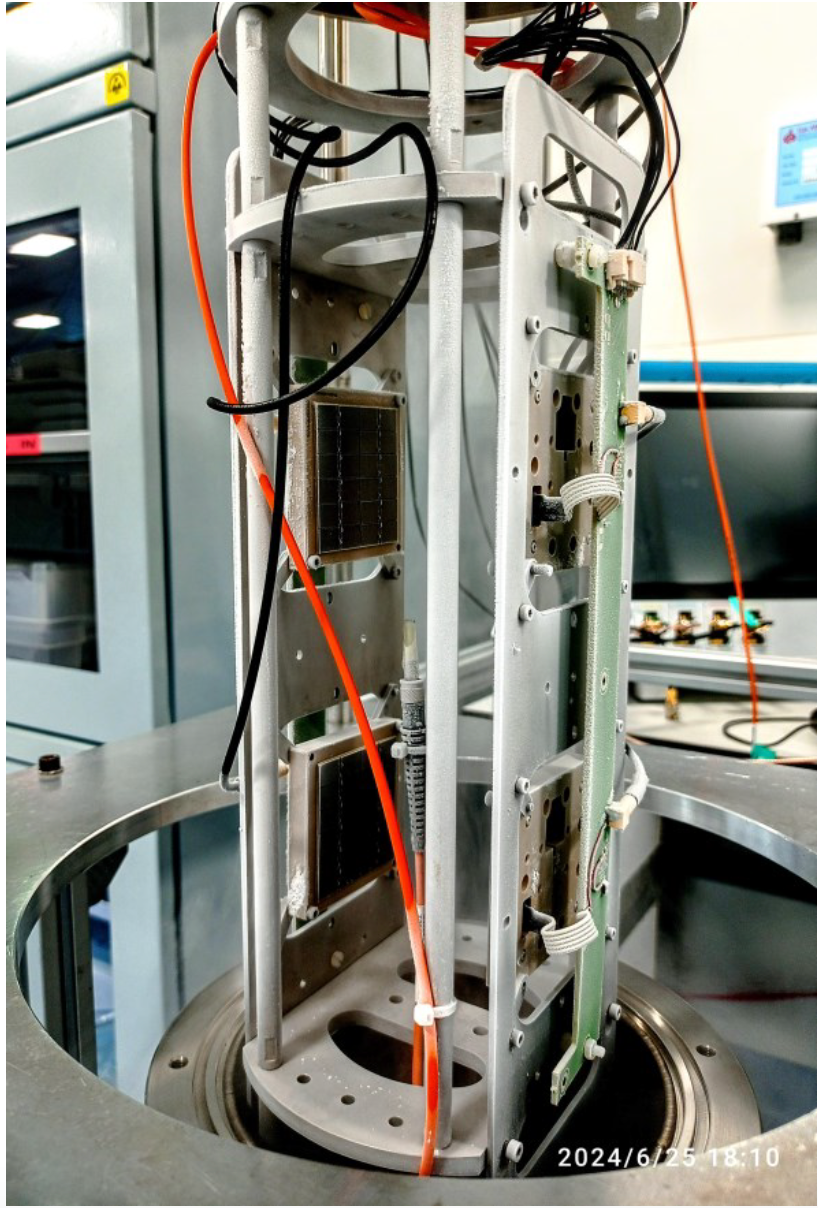}\label{fig:vTile-test-stand-a}}
    \qquad
    \subfloat[Cold test stand schematic]{\includegraphics[width=0.45\textwidth]{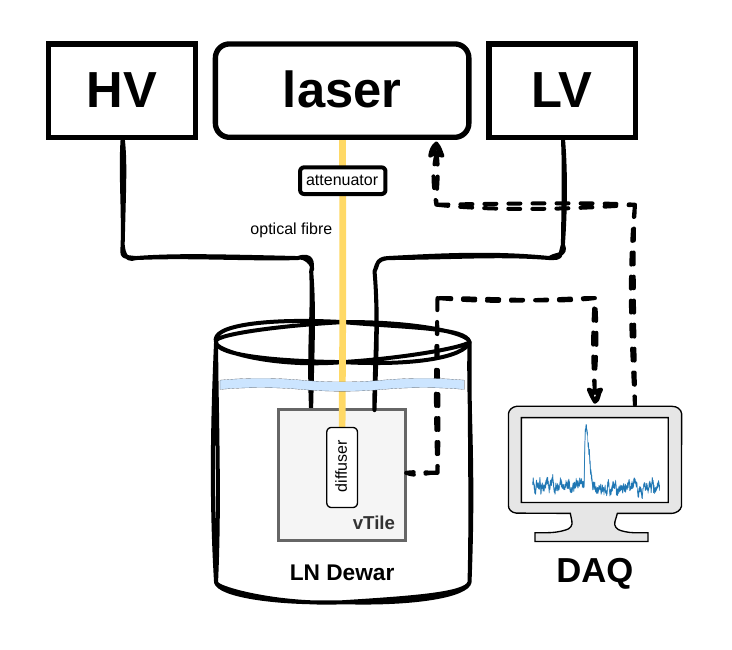}\label{fig:vTile-test-stand-b}}
    \caption{vTile test stand. \subfigref{fig:vTile-test-stand-a} Picture of 4 vTiles mounted on the testing carrier in the test stand, showing the optical fibre and the light diffuser facing the SiPMs. Up to 4 vTiles can be tested at the same time.
    \subfigref{fig:vTile-test-stand-b} Schematic of the test stand electronics and readout: high voltage power supply (HV) for the SiPM sensor bias, low voltage power supply (LV) to enable the ASIC amplifier, the laser light source and the DAQ controlling also the laser trigger.}
    \label{fig:vTile-test-stand}
\end{figure}

Identical custom-made test stands are operated at the University of Oxford and STFC Interconnect in ISO-7 cleanroom areas. Each test stand consists of a stainless steel frame, which can host up to 4 vTiles at the same time (Fig.~\ref{fig:vTile-test-stand-a}). The frame hosting vTiles can be raised and lowered into a dewar flask designed to hold 10\,L of liquid nitrogen. Each vTile is connected through a six-pin connector to a test PCB, used to power the ASIC, the analogue switch, to deliver the bias voltage, and to read out the vTile output signal. 
The test PCB is connected through feedthroughs in the dewar flange to two external commercial power supplies: a low voltage supply, which powers the ASIC (at 2.2\,V) and the analogue switch (at 2.5\,V), and a high voltage supply which can bias the vTile up to 75\,V. 
A 405\,nm class-3 laser is used as a pulsed light source to illuminate the SiPMs through an optical fibre connected to a plastic diffuser, centred with respect to the 4-vTile carrier.
Data acquisition of single vTile waveforms is performed using a Tektronix MS058 oscilloscope interfaced with a computer, triggered on the pulsed laser synchronised output (Fig.~\ref{fig:vTile-test-stand-b}).
Data is digitised at a sampling rate of 125\,MHz with a bandwidth of 250\,MHz.

On arrival at the cold test sites, vTiles are unbagged in the cleanroom and scanned to measure their particulate count as described in Sec.~\ref{sec:component-metrology}. vTiles are then loaded into the test stand steel frame, lowered into a closed volume with identical dimensions to the dewar flask, and tested at room temperature. This first test for each vTile measures its reverse-bias I-V curve, the RMS of the baseline voltage, and the noise spectrum while the vTile is biased at 20\,V to operate well below the expected breakdown voltage. The vTiles passing the room temperature test QA/QC criteria, summarised in Table~\ref{table:specs}, proceed to cryogenic testing. The steel frame assembly is lowered into the dewar flask filled with liquid nitrogen, and a light-tight cleanroom blanket is secured around the outside of the dewar flange.
The cryogenic test consists of an I-V measurement, acquiring noise power spectra with the vTile biased at 20\,V, and acquisition of waveforms for laser calibration at each vTile bias in 0.3\,V steps between 66\,V and 75\,V, corresponding approximately to overvoltages of 5\,V to 10\,V.

\paragraph{Breakdown voltage measurement}\label{sec:breakdown-measurement}

\begin{figure*}[htb!]
  \centering
  \subfloat[vTile I-V curve at room temperature.]{\includegraphics[width=0.35\textwidth]{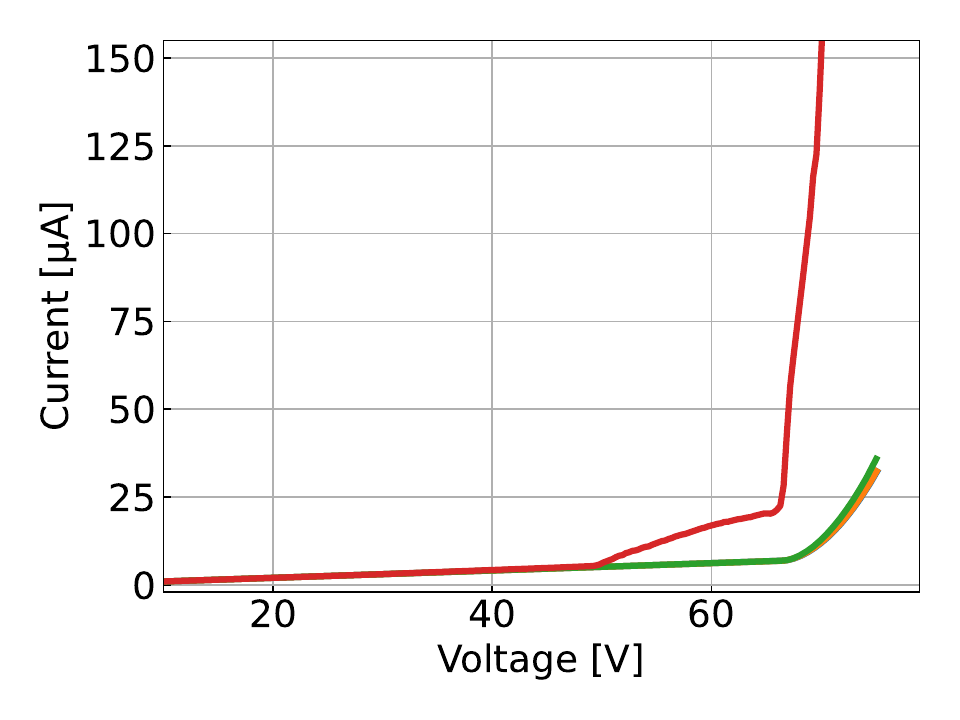}\label{fig:warmIV}} 
  \subfloat[vTile I-V curve at cryogenic temperature.]{\includegraphics[width=0.35\textwidth]{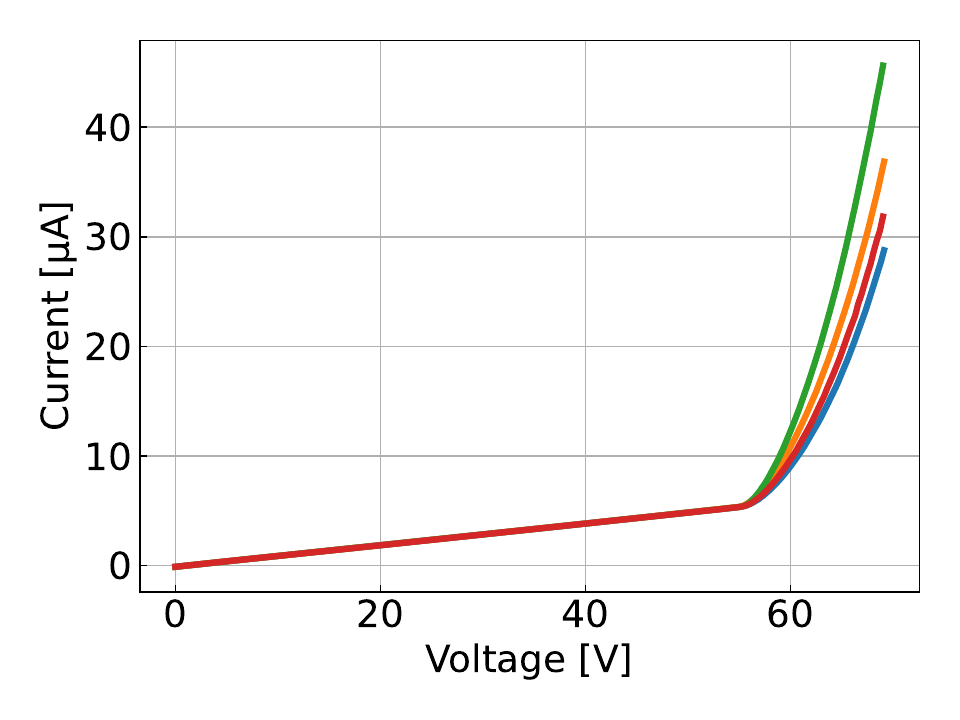}\label{fig:coldIV}} \qquad
  \subfloat[vTile cold breakdown voltage distribution \\before replacement of SiPMs.]{\includegraphics[width=0.35\textwidth]{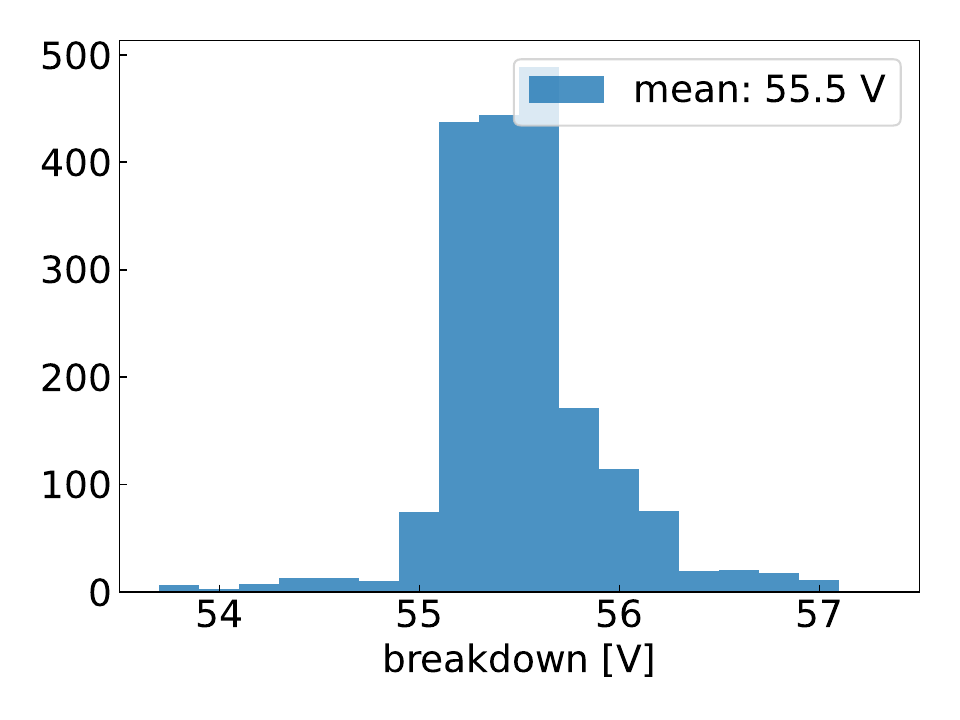}\label{fig:QAQC-breakdown}}
  \subfloat[single SiPM I-V curves at room temperature.]{\includegraphics[width=0.35\textwidth]{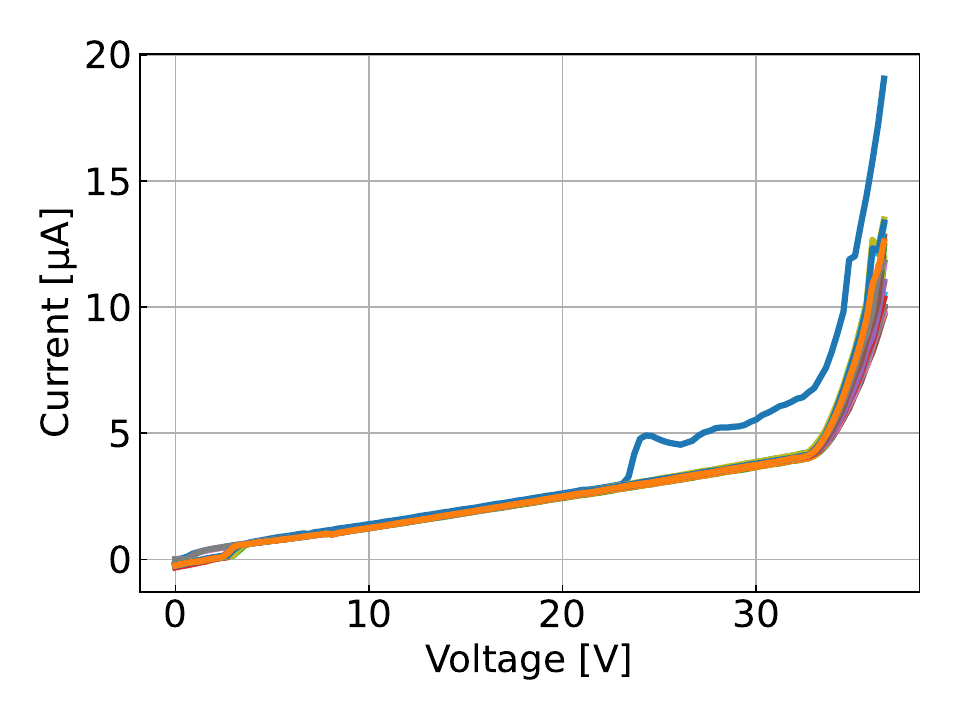}\label{fig:singleIV}}
  \caption{Example of reverse-bias I-V curves for multiple vTiles at warm \subfigref{fig:warmIV} and at cold \subfigref{fig:coldIV} temperatures. A warm I-V curve for a vTile (in red) shows a double breakdown, while all the others have a clear single breakdown at around 66\,V at warm and 55\,V at cold temperatures.
  \subfigref{fig:QAQC-breakdown} Breakdown distribution for production vTiles tested in cold. 
  \subfigref{fig:singleIV} Example of 24 single SiPM I-V curves for a vTile showing a double breakdown: most SiPMs show a breakdown at 32\,V, except one (blue I-V curve) that shows a double breakdown due to a SiPM defect.
  }\label{fig:IV}
\end{figure*}

To verify the vTile survived transport undamaged, the I-V curve is measured at room temperature in the dark dewar. The I-V is characterised by two zones: pre-breakdown, in which the current increases linearly with the voltage, and post-breakdown, which corresponds to the Geiger regime, where an avalanche of charge carriers can be initiated by single photons. 
Figure~\ref{fig:IV} shows examples of room temperature and liquid nitrogen temperature I-V curves.

The method employed to determine the breakdown voltage from an I-V measurement is the ``second logarithmic derivative''~\cite{IVfitting}.
This identifies the breakdown voltage as the voltage that corresponds to the maximum of the second derivative of the logarithm of the smoothed I-V curve. This method proved to be reliable for finding the transition point across I–V measurements of varying quality.
Due to the configuration of the SiPMs in a vTile (as described in Sec.~\ref{sec:vpdu}), the vTile breakdown voltage is twice the single SiPM breakdown (32\,V at warm and 27\,V at cold~\cite{DarkSide-20k:2024usz}), giving nominal values of 67\,V at room temperature (and 55\,V at liquid nitrogen temperature). vTiles proceed to subsequent testing if the breakdown is within~$\pm$1\,V of the nominal values. The distribution of the breakdown measured at cryogenic temperature for 2208 vTiles is shown in Fig.~\ref{fig:QAQC-breakdown}.
The difference between liquid nitrogen and liquid argon temperatures corresponds to a breakdown-voltage shift of only a few hundred millivolts, well below the operating overvoltage.
SiPMs have been selected to have a breakdown-voltage spread of $\pm$1\,V and are operated in a regime where such variations in the bias voltage result in only minimal changes to the photon detection efficiency, with a maximum variation of about 1.3 percentage points~\cite{sipm_gola}.
vTiles that show a ``double breakdown'', i.e.\ a double change in the slope of the I-V curve (example in the red I-V curve of Fig.~\ref{fig:warmIV}), are flagged at this stage and rejected from the production.
Double breakdown can be due to mechanical damage of a SiPM during production or transportation, or to a short circuit caused by a wire bond deformation such that it contacts a SiPM.
To identify the root cause of a double breakdown, an additional room temperature diagnostic test was developed in which each SiPM is individually biased in order to measure its I-V curve. In Fig.~\ref{fig:singleIV} single SiPM I-V tests for a vTile experiencing a double breakdown are reported, showing that most SiPMs have I-V curves as expected, except one (in blue) which shows an early change of slope around 20\,V. In this case, a repair procedure is enacted to replace the single identified problematic SiPM on the vTile. The procedure consists of removing the wire bonds and the damaged SiPM, cleaning the residual solder, replacing the SiPM with a new one, reflow, and re-wire-bonding. This rework process was successful in recovering double breakdown vTiles over 86\% of times.

\begin{figure}[htb!]
    \centering
    \includegraphics[width=0.49\textwidth, trim=45 10 50 20, clip]{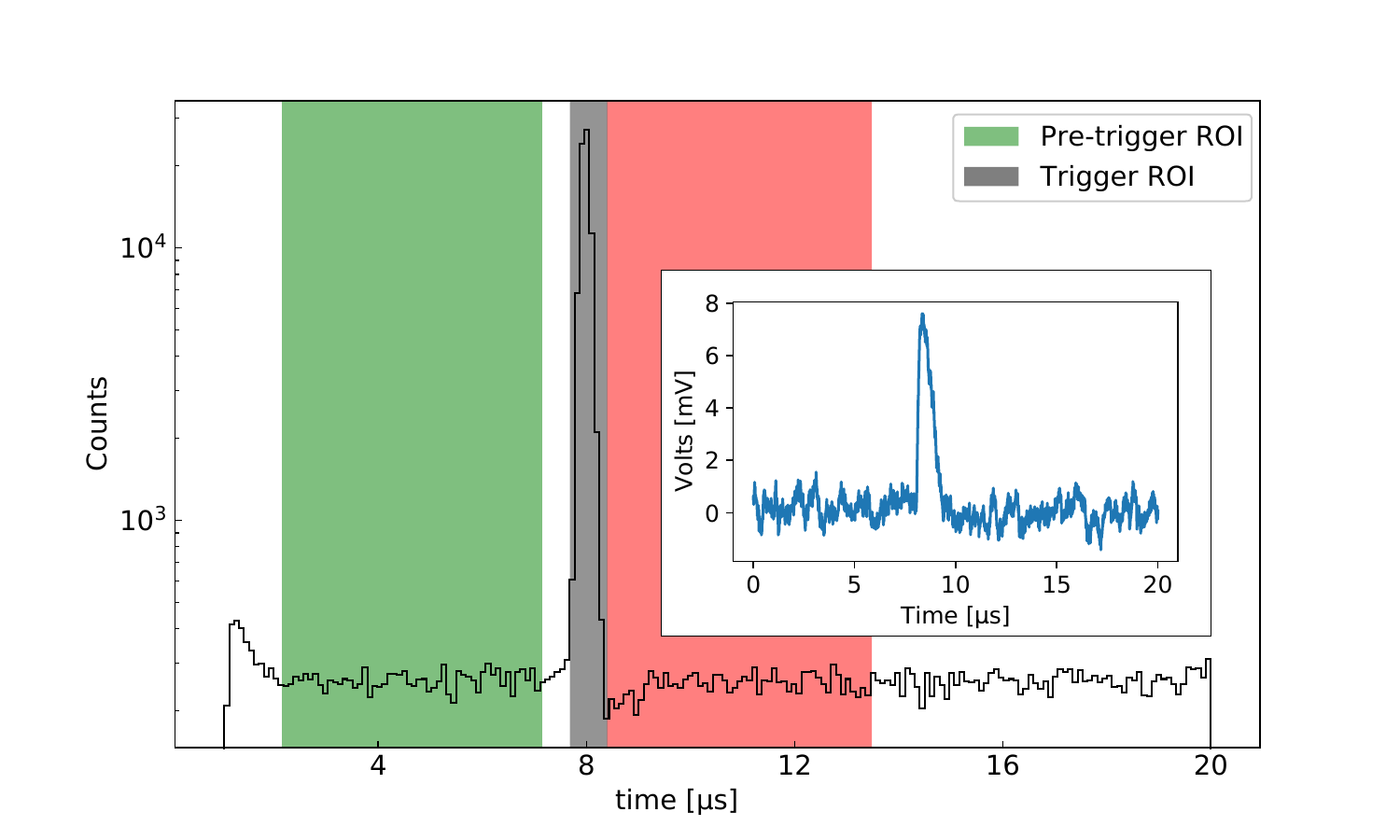}
    \caption{Time distribution of pulses, showing the trigger region (grey), the pre-trigger region (green) used to evaluate the uncorrelated noise and the post-trigger region (red). The inset shows a single waveform for a vTile biased at 69\,V.}
    \label{fig:passport}
\end{figure}

\paragraph{Signal-to-Noise ratio for single photoelectron detection}
\label{sec:SPE-resolution}
During the cryogenic tests, the response of each vTile to 405\,nm laser pulses is recorded in 3$\times$10$^5$ 20\,$\mu$s-long waveforms, while the vTile is biased between 66\,V and 75\,V in 1\,V steps. 
The waveforms are reconstructed offline with a peak finding algorithm which identifies a pulse if the amplitude is greater than 8 times the RMS of the baseline, where the RMS is calculated over a 1\,$\mu$s window at the beginning of the waveform, before the laser trigger region. Figure~\ref{fig:passport} shows the distribution of the times associated with the maxima of the pulses identified by the peak finding algorithm when a vTile is biased at 69\,V and an example of a waveform.

%
\begin{figure}[ht!]
    \centering
    \subfloat[Amplitude spectrum of a nominal vTile.]
    {\includegraphics[width=0.45\textwidth]{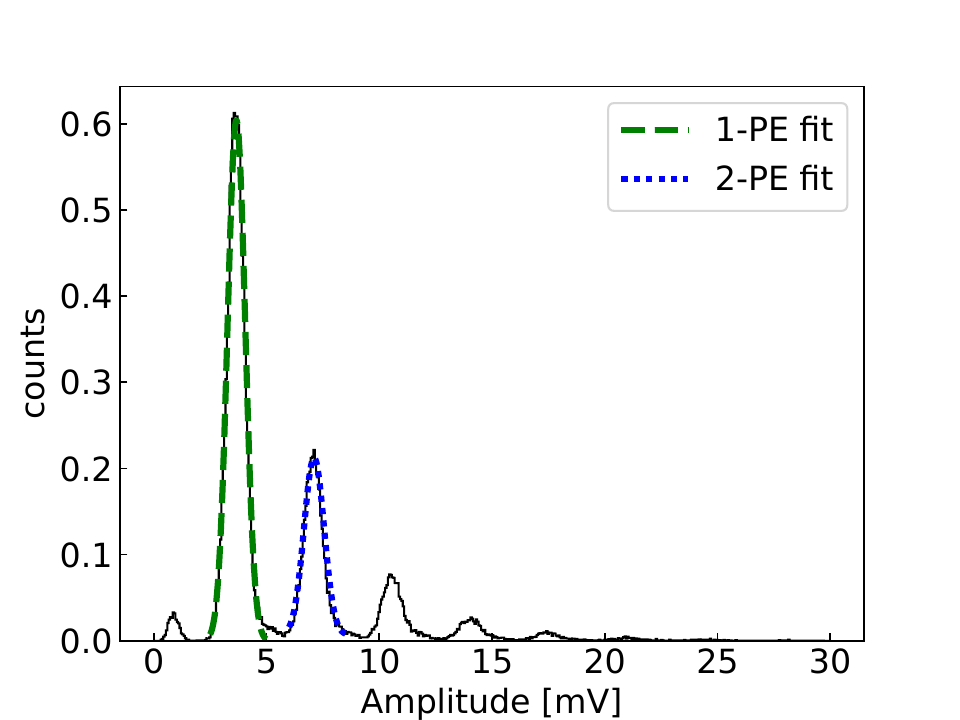}\label{fig:finger_plot}}
    \qquad
    \subfloat[SNR vs.\ vTile bias voltage.]
    {\includegraphics[width=0.47\textwidth]{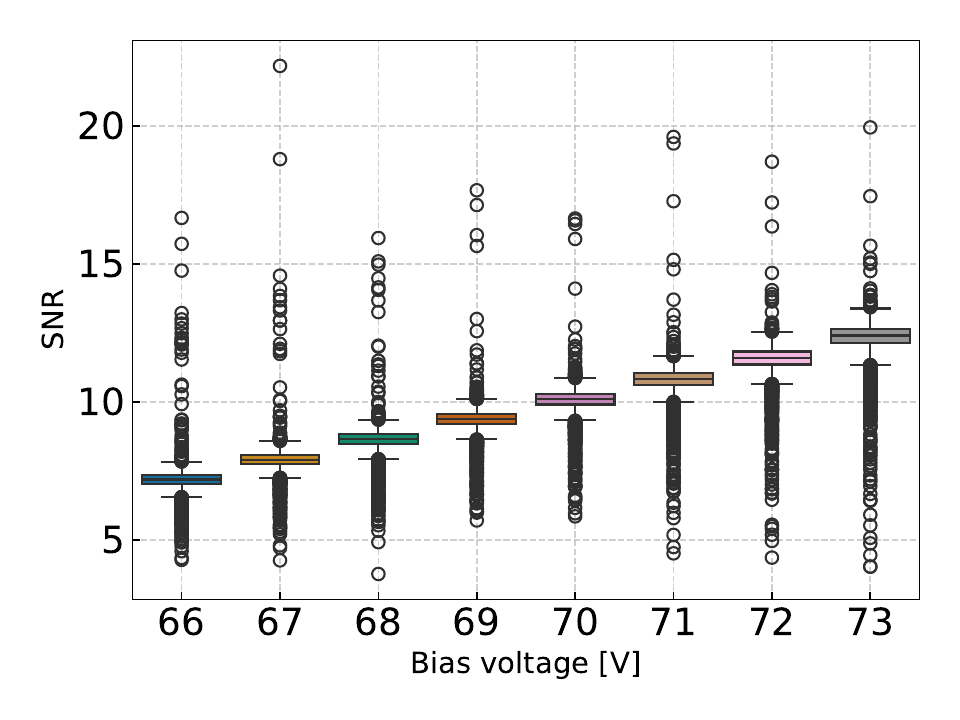}\label{fig:snr}}
    \caption{\subfigref{fig:finger_plot}
    Amplitude spectrum for a vTile biased at 69\,V together with a Gaussian fit to the 1 and 2-PE peaks (in green and blue respectively) used to compute the SNR. The leftmost small peak corresponds to the residual pedestal.
    \subfigref{fig:snr} SNR plotted against the vTile bias voltage for 2208 production vTiles showing a clear linear correlation with the bias voltage. Boxes represent data between the first and third quartiles. Dots represent outliers more than 1.5 interquartile range away from the top or bottom of each box.}
    \label{fig:QAQC1}
\end{figure}

\begin{figure}[ht]
    \centering
    \subfloat[]{\includegraphics[width=0.40\textwidth]{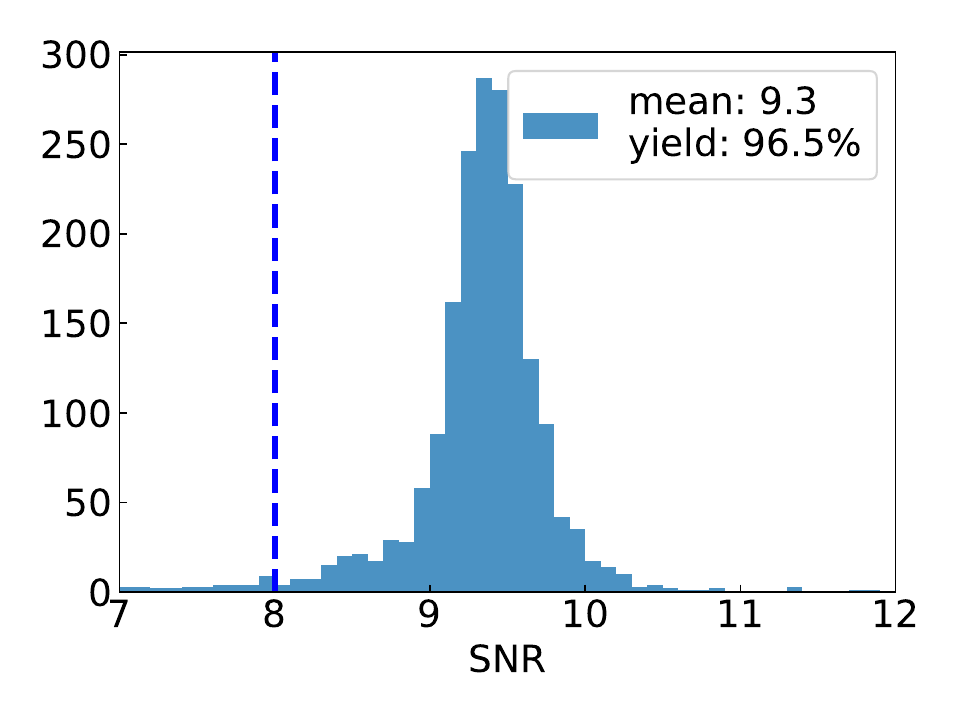}\label{fig:QAQC-snr}} \\
    \subfloat[]{\includegraphics[width=0.40\textwidth]{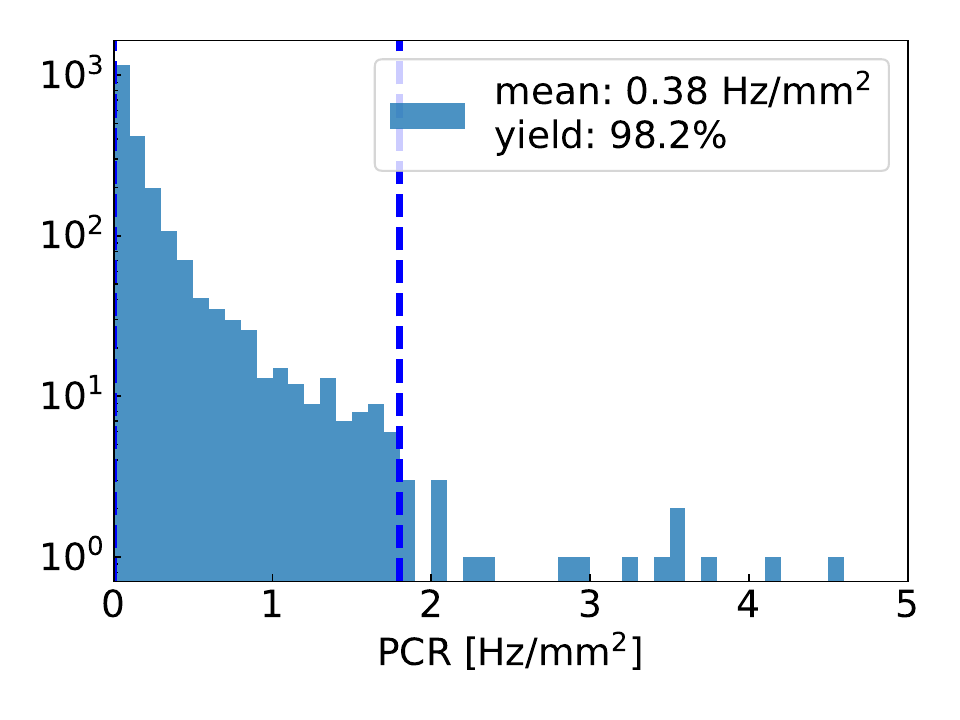}\label{fig:QAQC-dcr}
    }
    \caption{\subfigref{fig:QAQC-snr} SNR, and \subfigref{fig:QAQC-dcr} PCR distributions for production vTiles tested in cold, with a bias of 69\,V. The dashed-blue lines correspond to the acceptance criteria (Table~\ref{table:specs}) and the yields correspond to the fraction of vTiles passing the criteria.}
    \label{fig:QAQC2}
\end{figure}

The signal-to-noise ratio (SNR) of a vTile is defined for the production QA/QC criteria as the distance between the 2 and 1 photoelectron peaks in the amplitude spectrum defined in the trigger region of interest (ROI), relative to the noise:
\begin{equation}
    \mathrm{SNR} = \frac{\mu_{2\mathrm{PE}}-\mu_{1\mathrm{PE}}}{\mathrm{RMS}}.
\end{equation}
where $\mu_{1\mathrm{PE}}$ and $\mu_{2\mathrm{PE}}$ are the mean values of the Gaussian fits to the one (``1-PE amplitude'') and two (``2-PE amplitude'') PE peaks respectively, as shown in Fig.~\ref{fig:finger_plot}, and RMS refers to the pedestal.
The 1-PE amplitude is required to be consistent with the designed board gain of 3.9\,mV.
This target value was defined during the front-end board design, simulation and optimization phase to provide a uniform 1-PE response for both TPC and veto tiles, and was subsequently validated through board characterisation measurements.
vTiles with 1-PE amplitude between 3.4--4.8\,mV are accepted, accounting for gain fluctuations caused by variations in both the PCB electronics and the intrinsic SiPM breakdown voltage.
vTiles exhibiting 1-PE amplitudes on the tail of the distribution are rejected: this behaviour is typically due to the presence of sub-peaks which can skew the reconstructed 1-PE amplitude up to 20\%, affecting the single PE resolution.
Such effects can arise from non-uniform gains among the 24 SiPMs (confirmed by differences in single I-V test), issues in the bias-voltage divider (not all SiPMs receive the same bias), or possible damage to the ASIC. 
The RMS of the 1-PE distribution is targeted to lie in the range 300--450\,$\mu$V in order to guarantee a SNR greater than 8.
Figure~\ref{fig:QAQC-snr} shows the SNR for 2208 vTiles tested at cryogenic temperature.
These QA/QC requirements allow each vPDU quadrant to achieve a SNR greater than 4 (since at the vPDU level, the RMS of the quadrants is obtained as the quadrature sum of the RMS values of the 4 individual vTiles), which preliminary characterisations indicate is sufficient to guarantee the required neutron veto efficiency.
The average SNR response grows linearly with vTile bias voltage, as shown in Fig.~\ref{fig:snr}.

\paragraph{Uncorrelated noise characteristics}\label{sec:noise}
Uncorrelated noise comes from avalanches produced by processes internal to the SPADs, rather than in response to incident photons, termed Dark Count Rate (DCR). At room temperature, DCR is produced by thermionic emission in the sensor, and at cryogenic temperature DCR is dominated by field-enhanced tunnelling of charge carriers to the SiPM junction~\cite{8818357}. 
As the sensors view the liquid nitrogen volume, we measure the Pulse Count Rate (PCR), which while correlated with the DCR additionally includes light produced in the liquid nitrogen volume by cosmogenic and environmental radiation. The PCR is calculated as the mean number of hits in the 5\,$\mathrm{\mu s}$ pre-trigger ROI (as shown in Fig.~\ref{fig:passport}), normalised by the SiPMs' active area of the vTile (2218\,mm$^2$).
The distribution of the PCR of 2208 vTiles is shown in Fig.~\ref{fig:QAQC-dcr}.
Dedicated measurements have established that the PCR is 1--2 orders of magnitude larger than the intrinsic DCR. For QA/QC, we consider the uniformity of the PCR measurement, and exclude the tails by rejecting vTiles with PCR values above 1.8\,Hz/mm$^2$. This threshold is chosen since at this value we start observing a degraded SPE detection resolution, usually correlated with SiPMs defects (e.g.\ metal breaks as in Fig.~\ref{fig:sipm_inspection}), which leads to an increase in pulse count and baseline fluctuations.
The test environment was also not deemed suitable for characterising the correlated noise sources in SiPMs (such as direct crosstalk and afterpulsing), quantities that were not assessed in the QA/QC studies.

After the cryogenic testing of vTiles, devices passing QA/QC are shipped for integration into vPDUs. Considering the SiPM reworking procedure, 91\% of vTiles pass the QA/QC criteria summarised in Table~\ref{table:specs}.
Devices failing QA/QC exit the production process at this point for root cause investigation using the specialised diagnostic tests described above.


\subsection{Veto Photo-Detector Unit Integration} \label{sec:vpdu_assembly}

Upon arrival at the integration sites, vTiles and veto motherboard PCBs undergo acceptance metrology. 
The vTiles are unbagged in the ISO-7 cleanroom environment and scanned, as described in Sec.~\ref{sec:dust-control} for dust counting analysis and for traceability of the origin of any mechanical damage relative to the exit scan at the vTile test site. 

Prior to integrating 16 vTiles, the vMB PCBs are initially tested in NOA at room and at cryogenic temperatures.
Each vMB is powered with 7\,V and the current draw is measured, with expected values of 7--8\,mA at warm temperature (5\,mA at cold).
Each quadrant is separately enabled, measuring the current drawn which is expected to be in the 46--48\,mA range at warm (25--26\,mA at cold), while for all four quadrants the nominal current draw is 165--166\,mA at warm (105--106\,mA at cold). Voltages are measured across the vMB to confirm the values to be provided to the vTiles, in particular for the ASIC and for the vTile power switch (both at warm and at cold).
A dummy load (equivalent to one vTile's load) is mounted in each of the 16 slots to measure the current draw at room temperature for an enabled vTile at 7\,V, expecting 72\,mA, and for a vTile biased at 40\,V, expecting 3.896\,$\mu$A.
In addition, a noise power spectrum vs.\ frequency is measured for each quadrant and compared with a reference noise power spectrum.
The vMBs that do not pass these criteria (mostly due to issues in the mounting of the components) are fixed, recovering the vast majority of them (98\%).
The vMBs are shipped to the vPDU integration sites, where the acceptance testing repeats the room temperature measurements described here. The vMBs that pass these criteria progress to vPDU integration.

\begin{figure}[ht]
    \centering
    \includegraphics[width=0.35\textwidth]{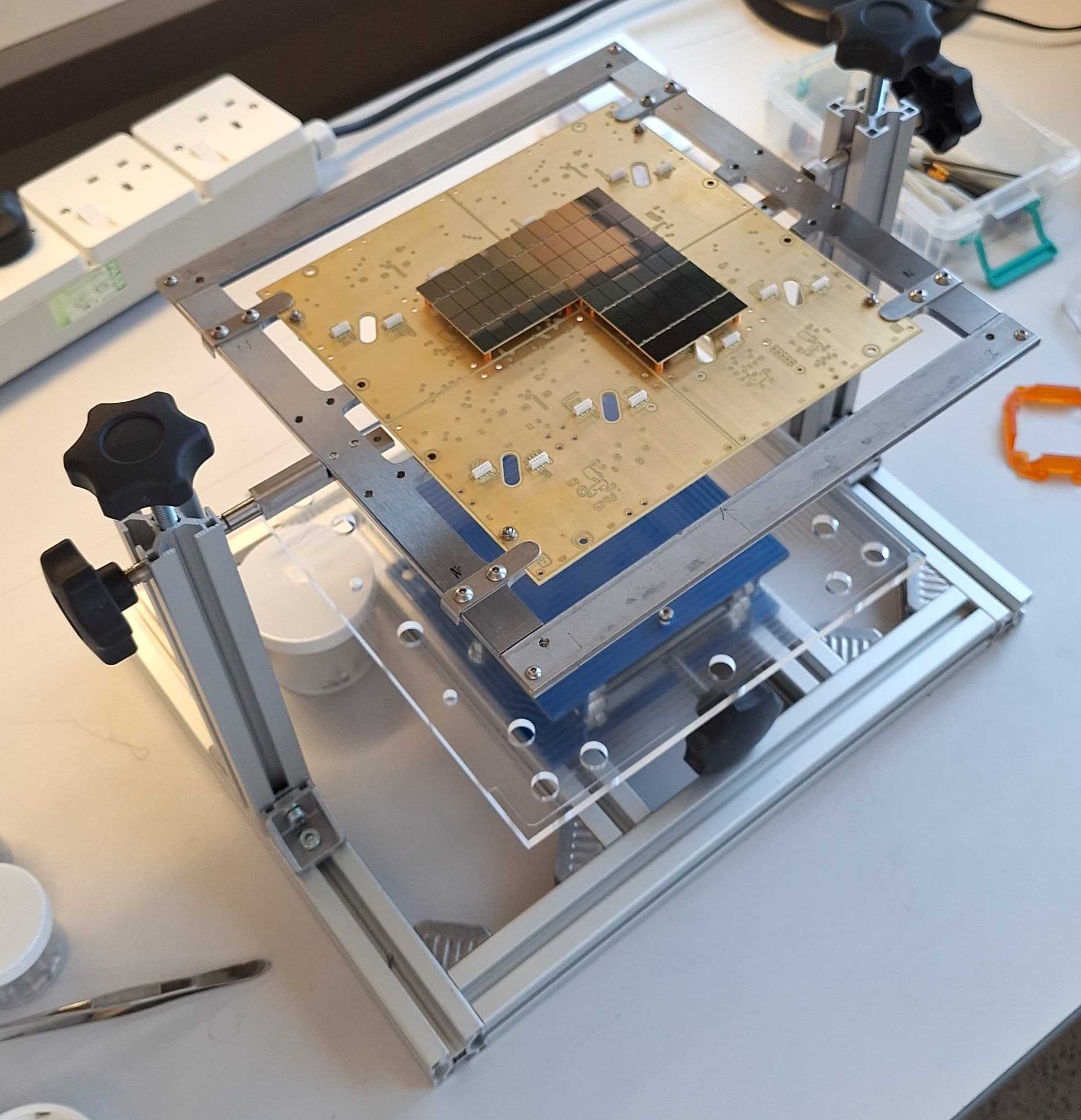}
    \caption{vPDU assembly: example of a vMB being populated with vTiles on an assembly fixture in Manchester.} \label{fig:vpdu-assembly-fixture}
\end{figure}

\paragraph{Integration of vTiles onto vMB}
Bespoke tooling is required to assemble vTiles onto the vMB, and the vMB into a custom handler fixture, shown in Fig.~\ref{fig:vpdu-assembly-fixture}.
The vMB is first fitted to a frame that can be rotated. This allows access to the top for vTile alignment and to the bottom for vTile fixing. 
Stainless steel rods with a threaded end are first used to remove vTiles from their transport carrier, contacting the vTiles only via their threaded copper pillars. Next, the rods are aligned through the corresponding holes in the vMB to allow accurate placement without vTiles contacting each other. Finally, the rods are lowered to mate the 6-pin electrical connector on the vTile to the corresponding socket on the vMB. The vTiles are fitted one by one to populate the entire vMB. Each vTile copper pillar is fixed by fitting with a stainless steel M1.6 cap head screw and nylon washer. Once the vMB is fully populated, it is removed from the assembly fixture and installed into the stainless-steel handler fixture assembly shown in Fig.~\ref{fig:vpdu}.
The vMB is held to the handler by 4 r-clips, which retain the stainless-steel pins of the vMB. The assembled vPDU is scanned for dust metrology, after which standoffs are installed through the handler steel, and an ESD-safe acrylic cover with slotted holes is installed above the vPDU to protect it for all subsequent testing and transport steps.

After integration, vPDUs are characterised at room temperature and at liquid nitrogen temperature in setups equivalent to those used for single vTile testing, to verify that the vTiles and vMBs were not damaged during the integration process and to confirm the photo detection performance of the completed objects.
Tests are performed on both individual vTiles as well as on full vPDU quadrants, extracting the same quantities that are used to evaluate the QA/QC of vTiles, as will be described in a forthcoming paper.

\subsection{Radiopurity control measures throughout the production process}

A key requirement for the vPDUs used to instrument the DarkSide-20k IV is that they meet stringent radioactivity and cleanliness specifications, enabling operation in an instrumental background-free regime for 10 years. Radiopurity requirements for each individual detector component are derived in order to keep the neutron-induced WIMP-like events surviving veto cuts -- potential backgrounds to the dark matter search -- to be an order of magnitude below the expected irreducible background from neutrino interaction on argon in a 200\,tonne-year exposure, expected to be 3.5\,events (updated from Refs.~\cite{Cadeddu2018_thesis, Cadeddu_2023}). In this context vPDUs should provide a negligible contribution to the neutron induced total background. In addition, the gamma rate induced by the photon detection system in the IV must be below 50\,Hz in order to minimise detector dead time.
Generic dust containing radioactive elements releases gammas and can induce neutron emission; dust is also expected to be the single largest contributor to radon emanation, therefore dust contamination on the wetted surfaces is required to be limited to less than 500\,ng/cm$^2$ (as for LZ~\cite{AKERIB2020163047}).

To ensure that materials used in the vPDUs meet this requirement, every vTile component has been assayed through three different screening procedures: Inductively Coupled Plasma Mass Spectrometry (ICP-MS), High Purity Germanium (HPGe) radiation detection and polonium extraction. These measurements give a clear picture of the radioactive budget of the uranium and thorium chains and gamma emitters, such as $^{40}$K. 
The principal source of neutron background comes from the lower part of the uranium chain from the capacitors, followed by the PCB (as its mass is dominant with respect to the other components). The full radioactive budget and mass for each component are reported in Table~\ref{tab:components}, while the total activities for all components resulting from these measurements are summarised in Table~\ref{tab:activities}.

\begin{table*}[ht]
\centering
\scriptsize
\caption{The radioactive budget for the total mass of each vTile component (e.g.\ 24 SiPMs, 4 copper pillars, 21 capacitors), based on assay results, is reported (activities in mBq). The dominant contribution comes from the lower part of the uranium chain of the capacitors and of the PCBs (which have the largest mass). The contribution of other components is subdominant.}
\begin{tabular}{c|c|c|c|c|c|c}
\hline
\textbf{Component} & \textbf{Mass} [mg] & \textbf{$^{238}$U-up} & \textbf{$^{238}$U-middle} & \textbf{$^{238}$U-bottom} & \textbf{$^{232}$Th}  & \textbf{$^{40}$K}  \\
\hline 
    SiPMs          & 2856 & 0.22 $\pm$ 0.16 & 0.14 $\pm$ 0.04 & 0.8 $\pm$ 0.6 & 0.022 $\pm$ 0.025 & $<$0.27 (68\% CL) \\
    resistors      & 111  & 0.314 $\pm$ 0.027 & 0.78 $\pm$ 0.08 & 1.69 $\pm$ 0.15 & 0.140 $\pm$ 0.011 & 0.18 $\pm$ 0.06 \\
    capacitors     & 662  & 0.068 $\pm$ 0.012 & 0.08$^{+0.04}_{-0.03}$ & 27.2 $\pm$ 1.7 & 0.12 $\pm$ 0.06 & 0.9$^{+0.4}_{-0.3}$\\
    solder (Indium Paste) & 20 & $<$4.2$\cdot$10$^{-5}$ (99.7\% CL) & $<$6.6$\cdot$10$^{-5}$ (99.7\% CL) & 0.276 $\pm$ 0.006 & $<$2.3$\cdot$10$^{-5}$ (99.7\% CL)& $<$4$\cdot$10$^{-3}$ (99.7\% CL) \\
    solder (CHIPQUIK) & 160 & $<$0.059 (95\% CL)& 0.00093 $\pm$ 0.00018 & 0.093 $\pm$ 0.018 & $<$1.6$\cdot$10$^{-3}$ (95\% CL)& 0.0023 $\pm$ 0.0006 \\
    PCB               & 4500 & 0.17 $\pm$ 0.05 & 0.157 $\pm$ 0.010 & 0.23 $\pm$ 0.07 & 0.153 $\pm$ 0.010 & 3.02 $\pm$ 0.24 \\
    ASIC           & 65   & $<$0.0084 (68\% CL) & 0.0153 $\pm$ 0.0030 & $<$0.040 (68\% CL) & $<$0.0016 (68\% CL)& 0.076 $\pm$ 0.021 \\
    switch         & 6.5  & 0.0009 $\pm$ 0.0005 & 0.00165 $\pm$ 0.00020 & 0.0022 $\pm$ 0.0009 & 0.00074 $\pm$ 0.00034 & 0.0095 $\pm$ 0.0023 \\ 
    copper pillars & 1406 & $<$0.11 (68\% CL) & 0.056 $\pm$ 0.011 & 2.9 $\pm$ 0.8 & 0.047$\pm$ 0.021 & 0.29 $\pm$ 0.16 \\ 
    connector      & 124  & 0.00186 $\pm$ 0.00027 & 0.0409 $\pm$ 0.0015 & 0.409 $\pm$ 0.015 & 0.0108 $\pm$ 0.0010 & 0.021 $\pm$ 0.005 \\ 
    \hline
    \end{tabular}
    \label{tab:components}
\end{table*}

\begin{table}[ht]
\centering
\caption{Comparison of activities from uranium and thorium chains and gamma emitters between summing all vTile components and assaying a fully populated vTile.}
\begin{tabular}{c|c|c}
\hline
\textbf{Isotope} & \textbf{Summed components} & \textbf{Populated vTile} \\
 & \textbf{activity} [mBq/vTile] & \textbf{activity} [mBq/vTile] \\
\hline 
    $^{238}$U-up      &  0.91  $^{+0.18}_{-0.17}$ & 1.2  $\pm$ 0.6\\
    $^{238}$U-middle  &  1.28  $\pm$ 0.10 & 0.9  $\pm$ 0.1 \\
    $^{238}$U-bottom  &  33.7  $\pm$ 1.9 & 60.2 $\pm$ 4.2 \\
    $^{232}$Th        &  0.50  $\pm$ 0.06 & 0.4  $\pm$ 0.2  \\
    $^{40}$K          &  4.8   $^{+0.5}_{-0.4}$ & 5.1  $\pm$ 1.0 \\
    \hline
    \end{tabular}
    \label{tab:activities}
\end{table}

The radioactive contamination does not come only from the material itself but can be introduced during the assembly and characterisation processes. In particular, the most prominent sources of contamination are introduced from the environment: $^{222}$Rn from the $^{238}$U chain and $^{220}$Rn from the $^{232}$Th chain.

Radon is gaseous so its daughters can be easily directly deposited in materials and also through dust. 
All production phases are performed in ISO-5--ISO-7 cleanrooms to control the presence of airborne particles and prevent any re-contamination. The cleanrooms have all been assessed with a RAD7 detector, resulting in a radon level typically of 2\,Bq/m$^3$, compatible with the cleanroom specifications required for DarkSide-20k. 

To verify the production process contribution, one fully populated vTile has been radio-assayed.
The comparison between an estimated activity of a vTile coming from summing all the individual components and from the assay of the fully assembled vTile is reported in Table~\ref{tab:activities}. The first column reports the sum of each component listed in Table~\ref{tab:components} and its associated uncertainties, taking into account that some values are direct measurements while others are upper limits.
The contributions of central values, estimated values, and upper limits of activities are accounted for by means of independent Monte Carlo simulations, which are summed a posteriori.
The second column reports the direct measurement and its direct error on the measurement of the populated board.
The radio-assay of fully populated vTiles is in general in agreement with the estimate based on individual components, resulting in a negligible additional effect to the neutron background and to the gamma rate coming from the assembly process, therefore validating the production line.

Based on these measurements, 120 vPDUs installed in the IV of DarkSide-20k yield a surviving neutron fraction of 5.6$\times$10$^{-3}$,
and a gamma rate in the IV of 25\,Hz.

The vPDUs would contribute approximately 4\% of the total neutron background, 20\% of the gamma rate in the IV, and a negligible gamma rate in the TPC, thus meeting the current expectation.

\subsubsection{Dust control}\label{sec:dust-control}
\begin{figure*}[htb!]
    \centering
    \includegraphics[width=0.9\linewidth]{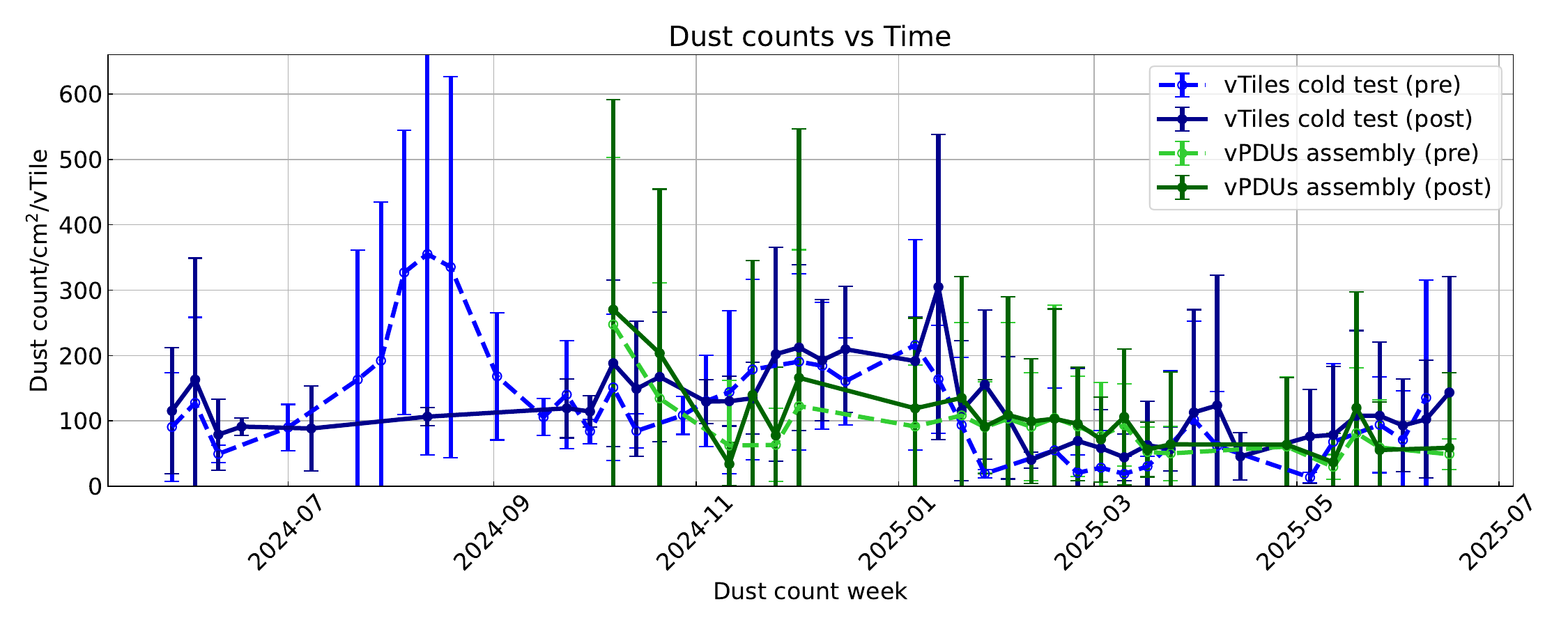}
    \caption{Distribution of the dust counts performed on the SiPM surface of vTiles in a cold test site (pre and post testing procedure) and in a vPDU assembly site (pre and post integration of vTiles onto vMBs). A constant amount of dust is measured, showing that no additional contamination is introduced during the testing and the production phases.}
    \label{fig:dust}
\end{figure*}

To monitor the radioactive contribution of production continuously throughout the process, each vTile and vPDU undergoes a comprehensive particulate analysis to ensure the integrity and cleanliness of its components.
The scanner data is used to perform a dust analysis on each vTile before and after cryogenic testing and before and after the vPDU integration.
Each vTile is scanned at a resolution of 6400\,DPI in 24-bit colour.
After scanning, a dust analysis is performed on each image. First, the image is converted to a 256 level greyscale image. This conversion simplifies the analysis by reducing the number of colour channels. To identify foreign particulates on the silicon surface, the image is thresholded to create a binary image, with particulates appearing as white and the background as black.
Subsequently, a particle analysis is conducted. Continuous white regions are detected using an 8-connectivity algorithm that identifies neighbouring white pixels, tracing and marking each region as a particulate. The algorithm continues searching for new white pixels until all white regions have been identified. Once complete, the number of particulates, their areas, circularity, aspect ratios, and coordinates are calculated and stored, in order to monitor the dust concentration during the production and testing phases of the sensors and over time in a given location as shown in Fig.~\ref{fig:dust}. The measured dust level has proven to be constant over the testing, production and integration phases, not exceeding on average 200\,counts/cm$^2$, or 400\,ng/cm$^2$ (assuming a spherical shape of the dust particles with a density of approximately 1\,g/cm$^3$ and a diameter above 10\,$\mu$m). Devices exceeding this typical value are investigated, and identified to be dicing residue permanently attached to the SiPM dies. It has been verified via radio-assay that these devices do not exceed the radioactivity of SiPMs with nominal surface quality. 

When not under active steps of the production process or triple-bagged for transport, PCBs, vTiles and vPDUs are stored in cabinets purged with ultra-high purity nitrogen (99.9998\%) in order to prevent any recontamination.

\subsection{Production yields}
\label{sec:specs}

\begin{figure}[htb!]
    \centering
    \includegraphics[width=0.99\linewidth]{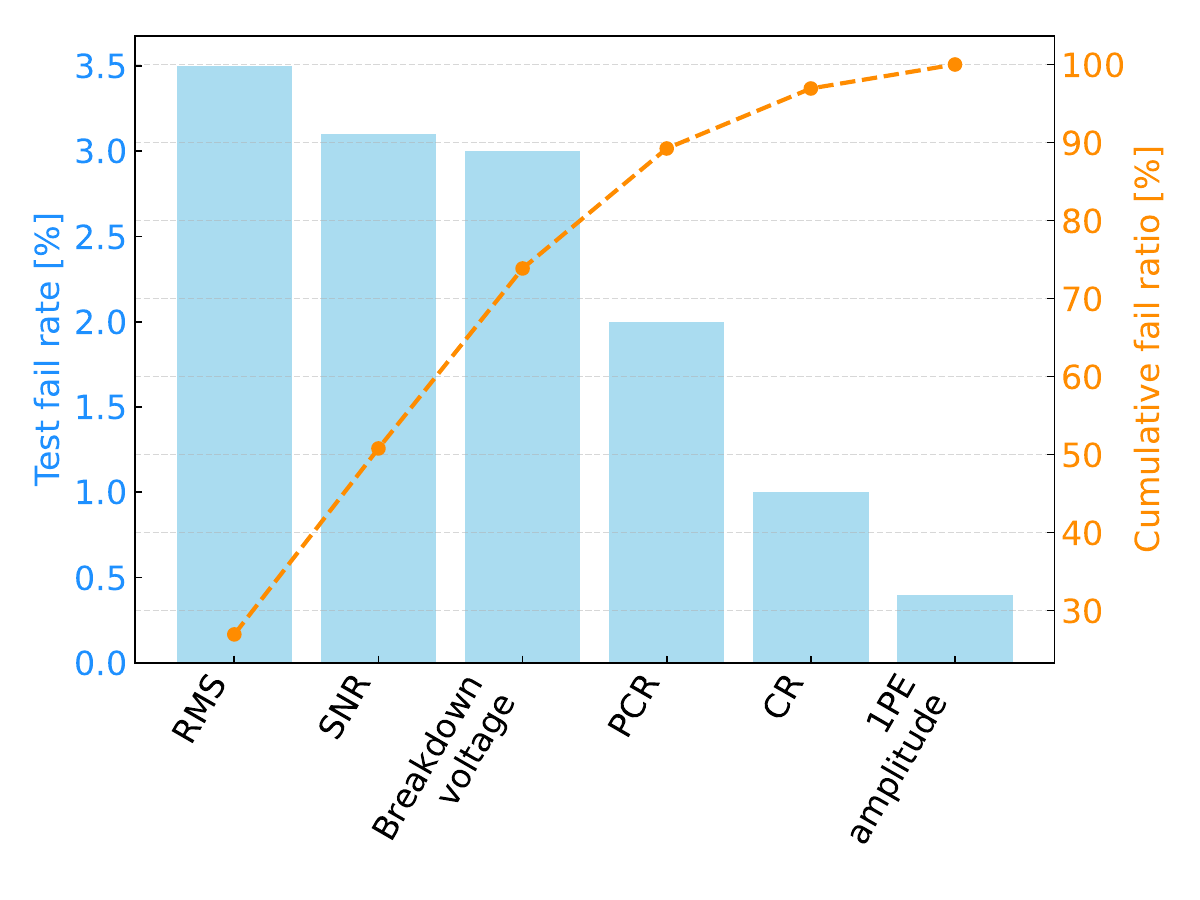}
    \caption{Pareto chart illustrating the most common failure modes in the vTile quality assessment process for tests performed at warm (CR, breakdown voltage from I-V curves) and at cold (RMS, 1-PE amplitude, SNR and PCR); all the failure modes are shown as independent rates, while vTiles can fail for multiple causes.
    Components or vTiles failing other earlier tests (as in Table~\ref{table:specs}) are not used in the production, therefore SiPMs are not wasted.}
    \label{fig:pareto}
\end{figure}

As described above, each vPDU component is characterised during every production step at room temperature and cryogenic conditions, to ensure conformity with the targeted specifications summarised in Table~\ref{table:specs}.
A summary of the most common failure modes in vTile testing is illustrated in the Pareto chart in Fig.~\ref{fig:pareto}, including only tests for which SiPMs would be wasted if the vTiles are rejected from production, whereas earlier tests (CI tests of ASICs and of unpopulated vTiles) would not cause wastage of SiPMs.
The total yield of the vTile production steps (CR, warm and cold exclusive tests) corresponds to 87\%.
For vTiles operating at their nominal breakdown voltage, the predominant failure modes stem from high RMS values and low SNR, which are typically correlated. PCR provides an additional means of identifying noisy vTiles, although it is not always an independent indicator.

\section{Conclusions}\label{sec:conc}

SiPM-based detectors constitute a technological innovation for future dark matter experiments. DarkSide-20k will implement the largest SiPM-based detector ever built, running for a decade starting from 2029 in order to push the boundaries for dark matter investigations.
The DarkSide-20k collaboration has assembled and characterised the total number of requested Veto Tiles necessary for populating 120 Veto Photo-Detector Units required to instrument the detector's Inner Veto.

Each component, from the single SiPM up to the final vPDU, underwent specific QA/QC steps in Italy, the United Kingdom, and Poland.
Most components that failed tests at early assembly stages were repaired, helping to maintain a high production yield and minimising the mortality of fully assembled units.

In particular, the total 87\% yield of the vTile production steps exceeds the 80\% specification of the DarkSide-20k production plan and ensures the completion of enough vPDUs to instrument the DarkSide-20k Inner Veto.

\begin{acknowledgements}
\input{acknowledgements-DS20k-2026-v4}
\end{acknowledgements}

\bibliographystyle{unsrt}
\bibliography{references}

\clearpage
\onecolumn
\section*{DarkSide-20k Collaboration}
\input{authors.tex}

\end{document}

%% file: acknowledgements-DS20k-2026-v4.tex
This work was supported by the U.S. National Science Foundation
(NSF) through Grants No. PHY-0919363, PHY-1004054, PHY-1004072,
PHY-1242585, PHY-1314483, PHY-1314507, PHY-1622337, PHY-1812482,
PHY-1812547, PHY-2310091, PHY-2310046, associated collaborative
grants No. PHY-1211308, PHY-1314501, PHY-1455351 and PHY-1606912, as
well as Major Research Instrumentation Grant No. MRI-1429544.
Additional support was provided by the Pacific Northwest National
Laboratory, operated by Battelle for the U.S. Department of Energy
under Contract No. DE-AC05-76RL01830.

Support was provided by the Istituto Nazionale di Fisica Nucleare
(INFN), through grants from the Italian Ministero dell'Istruzione,
Università e Ricerca, including Progetto Premiale 2013 and
Commissione Scientifica Nazionale II, as well as by the PRIN2020
project of the Italian Ministry of Research (MUR) (Grant No. PRIN
20208XN9TZ).

This work was supported by Canada Foundation for Innovation (CFI),
the Natural Sciences and Engineering
Research Council of Canada, SNOLAB, and the Arthur B. McDonald
Canadian Astroparticle Physics Research Institute.

Support was received from the French government from LabEx
UnivEarthS (ANR-10-LABX-0023 and ANR-18-IDEX-0001). Additional
support was received from the IN2P3-COPIN consortium (Grant No.
20-152).

This work was supported by the Chinese Academy of Sciences
(113111KYSB20210030) and the National Natural Science Foundation of
China (12020101004).
Support was provided by the São Paulo Research Foundation (FAPESP)
under Grant No. 2021/11489-7 and by the National Council for
Scientific and Technological Development (CNPq).
Support is acknowledged from the Deutsche Forschungsgemeinschaft
(DFG, German Research Foundation) under Germany's Excellence
Strategy -- EXC 2121: Quantum Universe -- 390833306.

The authors acknowledge support from the Spanish Ministry of Science
and Innovation (MICINN) through Grants PID2022-138357NB-C22 and
PID2022-138357NB-C21 and the Atracción de Talento Grant 2018-T2/
TIC-10494.

This work was supported by the Polish National Science Centre (NCN)
through Grants No. UMO-2022/47/B/ST2/02015, UMO-2023/51/B/ST2/02099 ,
and UMO-2023/50/A/ST2/00651, 
by the Polish Ministry of Science and Higher Education
(MNiSW, Grant No. 6811/IA/SP/2018).

This work was supported by the FNP IRA programmes: AstroCeNT
(MAB/2018/7), funded from the ERDF, and Astrocent
(FENG.02.01-IP.05-A015/25) co-financed by the European Union under
FENG 2021–2027; and Teaming for Excellence grant Astrocent Plus
(101137080)
funded by the European Union with complementary national funding from
the MNiSW (MNiSW/2025/DIR/811).

This project received funding from the European Union's Horizon 2020
research and innovation programme under Grant Agreement No. 952480
(DarkWave).

Support was provided by the Science and Technology Facilities
Council, part of United Kingdom Research and Innovation, and by The
Royal Society.

%% file: authors.tex
\setlength{\parindent}{0pt}

Fabio Acerbi\textsuperscript{\hyperlink{aff:1}{1}}, Pushparaj Adhikari\textsuperscript{\hyperlink{aff:2}{2}}, Paolo Agnes\textsuperscript{\hyperlink{aff:3}{3},\hyperlink{aff:4}{4}}, Iftikhar Ahmad\textsuperscript{\hyperlink{aff:5}{5}}, Sebastiano Albergo\textsuperscript{\hyperlink{aff:6}{6},\hyperlink{aff:7}{7}}, Ivone F.M. Albuquerque\textsuperscript{\hyperlink{aff:8}{8}}, Thomas Alexander\textsuperscript{\hyperlink{aff:9}{9}}, Andrew Knight Alton\textsuperscript{\hyperlink{aff:10}{10}}, Pierre-Andr\'e Amaudruz\textsuperscript{\hyperlink{aff:11}{11}}, Gioacchino Alex Anastasi\textsuperscript{\hyperlink{aff:6}{6},\hyperlink{aff:7}{7}}
% ...

\vspace{6mm}

\hypertarget{aff:1}{\textsuperscript{1}} Fondazione Bruno Kessler, Povo, 38123, Italy\\
\hypertarget{aff:2}{\textsuperscript{2}} Department of Physics, Carleton University, Ottawa, ON K1S 5B6, Canada\\
\hypertarget{aff:3}{\textsuperscript{3}} Gran Sasso Science Institute, L'Aquila, 67100, Italy\\
\hypertarget{aff:4}{\textsuperscript{4}} INFN Laboratori Nazionali del Gran Sasso, Assergi (AQ), 67100, Italy\\
\hypertarget{aff:5}{\textsuperscript{5}} Center for Experimental Cosmology \& Instrumentation, Department of Physics and Astronomy, University of California, Riverside, CA 92507, USA\\
\hypertarget{aff:6}{\textsuperscript{6}} INFN Catania, Catania, 95121, Italy\\
\hypertarget{aff:7}{\textsuperscript{7}} Universit\`a of Catania, Catania, 95124, Italy\\
\hypertarget{aff:8}{\textsuperscript{8}} Instituto de F\'isica, Universidade de S\~ao Paulo, S\~ao Paulo, 05508-090, Brazil\\
\hypertarget{aff:9}{\textsuperscript{9}} Pacific Northwest National Laboratory, Richland, WA 99352, USA\\
\hypertarget{aff:10}{\textsuperscript{10}} Physics Department, Augustana University, Sioux Falls, SD 57197, USA\\
\hypertarget{aff:11}{\textsuperscript{11}} TRIUMF, 4004 Wesbrook Mall, Vancouver, BC V6T 2A3, Canada\\
\hypertarget{aff:12}{\textsuperscript{12}} Physics Department, Columbia University, New York, NY 10027, USA\\
\hypertarget{aff:13}{\textsuperscript{13}} Department of Physics, University of Alberta, Edmonton, AB T6G 2R3, Canada\\
\hypertarget{aff:14}{\textsuperscript{14}} INFN Laboratori Nazionali di Legnaro, Legnaro (Padova), 35020, Italy\\
\hypertarget{aff:15}{\textsuperscript{15}} Savannah River National Laboratory, Aiken, SC 29808, United States\\
\hypertarget{aff:16}{\textsuperscript{16}} Institute for Experimental Physics, University of Hamburg, Hamburg, 22761, Germany\\
\hypertarget{aff:17}{\textsuperscript{17}} CIEMAT, Centro de Investigaciones Energ\'eticas, Medioambientales y Tecnol\'ogicas, Madrid, 28040, Spain\\
\hypertarget{aff:18}{\textsuperscript{18}} Centre de Physique des Particules de Marseille, Aix Marseille Univ, CNRS/IN2P3, CPPM, Marseille, France\\
\hypertarget{aff:19}{\textsuperscript{19}} INFN Pisa, Pisa, 56127, Italy\\
\hypertarget{aff:20}{\textsuperscript{20}} Physics Department, Universit\`a degli Studi di Pisa, Pisa, 56127, Italy\\
\hypertarget{aff:21}{\textsuperscript{21}} Williams College, Department of Physics and Astronomy, Williamstown, MA 01267, USA\\
\hypertarget{aff:22}{\textsuperscript{22}} Centro de Astropart\'iculas y F\'isica de Altas Energ\'ias, Universidad de Zaragoza, Zaragoza, 50009, Spain\\
\hypertarget{aff:23}{\textsuperscript{23}} Department of Physics, University of Oxford, Oxford, OX1 3RH, UK\\
\hypertarget{aff:24}{\textsuperscript{24}} INFN Torino, Torino, 10125, Italy\\
\hypertarget{aff:25}{\textsuperscript{25}} Department of Electronics and Communications, Politecnico di Torino, Torino, 10129, Italy\\
\hypertarget{aff:26}{\textsuperscript{26}} INFN Sezione di Roma, Roma, 00185, Italy\\
\hypertarget{aff:27}{\textsuperscript{27}} INFN Cagliari, Cagliari, 09042, Italy\\
\hypertarget{aff:28}{\textsuperscript{28}} Physics Department, Universit\`a degli Studi di Genova, Genova, 16146, Italy\\
\hypertarget{aff:29}{\textsuperscript{29}} INFN Genova, Genova, 16146, Italy\\
\hypertarget{aff:30}{\textsuperscript{30}} Warsaw University of Technology, Warsaw, 00-661, Poland\\
\hypertarget{aff:31}{\textsuperscript{31}} INFN Roma Tre, Roma, 00146, Italy\\
\hypertarget{aff:32}{\textsuperscript{32}} Mathematics and Physics Department, Universit\`a degli Studi Roma Tre, Roma, 00146, Italy\\
\hypertarget{aff:33}{\textsuperscript{33}} INFN Napoli, Napoli, 80126, Italy\\
\hypertarget{aff:34}{\textsuperscript{34}} Physics Department, Universit\`a degli Studi ``Federico II'' di Napoli, Napoli, 80126, Italy\\
\hypertarget{aff:35}{\textsuperscript{35}} Virginia Tech, Blacksburg, VA 24061, USA\\
\hypertarget{aff:36}{\textsuperscript{36}} ORCID 0000-0002-1767-1754, ORCID 0000-0002-4351-2255, ORCID 0000-0002-6394-9219, ORCID 0000-0002-8468-9540, ORCID 0000-0003-2869-2363, ORCID 0009-0009-0770-8830, ORCID 0000-0002-0597-2234, ORCID 0000-0001-5532-7711, ORCID 0000-0002-1455-4341, ORCID 0000-0003-2141-8230, ORCID 0009-0005-0286-0156, ORCID 0000-0001-5128-9279, ORCID 0000-0002-5527-4880, ORCID 0000-0002-5238-0442, ORCID 0000-0003-2952-6412, ORCID 0000-0002-2624-9416\\
\hypertarget{aff:37}{\textsuperscript{37}} AstroCeNT, Nicolaus Copernicus Astronomical Center of the Polish Academy of Sciences, Warsaw, 00-614, Poland\\
\hypertarget{aff:38}{\textsuperscript{38}} Department of Physics and Astronomy, Universit\`a degli Studi di Bologna, Bologna, 40126, Italy\\
\hypertarget{aff:39}{\textsuperscript{39}} INFN Bologna, Bologna, 40126, Italy\\
\hypertarget{aff:40}{\textsuperscript{40}} M. Smoluchowski Institute of Physics, Jagiellonian University, Krakow, 30-348, Poland\\
\hypertarget{aff:41}{\textsuperscript{41}} Physics Department, Universit\`a degli Studi di Milano, Milano, 20133, Italy\\
\hypertarget{aff:42}{\textsuperscript{42}} INFN Milano, Milano, 20133, Italy\\
\hypertarget{aff:43}{\textsuperscript{43}} Science \& Technology Facilities Council (STFC), Rutherford Appleton Laboratory, Technology, Harwell Oxford, Didcot, OX11 0QX, UK\\
\hypertarget{aff:44}{\textsuperscript{44}} Physics Department, Sapienza Universit\`a di Roma, Roma, 00185, Italy\\
\hypertarget{aff:45}{\textsuperscript{45}} Department of Physics, Engineering Physics and Astronomy, Queen's University, Kingston, ON K7L 3N6, Canada\\
\hypertarget{aff:46}{\textsuperscript{46}} Department of Physics, University of California Davis, Davis, CA 95616, USA\\
\hypertarget{aff:47}{\textsuperscript{47}} School of Physics and Astronomy, University of Edinburgh, Edinburgh, EH9 3FD, UK\\
\hypertarget{aff:48}{\textsuperscript{48}} Physics Department, Universit\`a degli Studi di Cagliari, Cagliari, 09042, Italy\\
\hypertarget{aff:49}{\textsuperscript{49}} APC, Universit\'e de Paris, CNRS, Astroparticule et Cosmologie, Paris, F-75013, France\\
\hypertarget{aff:50}{\textsuperscript{50}} Department of Physics, University of Liverpool, The Oliver Lodge Laboratory, Liverpool, L69 7ZE, UK\\
\hypertarget{aff:51}{\textsuperscript{51}} Physics Department, Princeton University, Princeton, NJ 08544, USA\\
\hypertarget{aff:52}{\textsuperscript{52}} Museo Storico della Fisica e Centro Studi e Ricerche Enrico Fermi, Roma, 00184, Italy\\
\hypertarget{aff:53}{\textsuperscript{53}} Department of Physics and Astronomy, The University of Manchester, Manchester, M13 9PL, UK\\
\hypertarget{aff:54}{\textsuperscript{54}} Department of Physics and Kavli Institute for Cosmological Physics, University of Chicago, Chicago, IL 60637, USA\\
\hypertarget{aff:55}{\textsuperscript{55}} INFN Laboratori Nazionali del Sud, Catania, 95123, Italy\\
\hypertarget{aff:56}{\textsuperscript{56}} Engineering and Architecture Department, Universit\`a di Enna Kore, Enna, 94100, Italy\\
\hypertarget{aff:57}{\textsuperscript{57}} Department of Physics and Astronomy, University of Hawai'i, Honolulu, HI 96822, USA\\
\hypertarget{aff:58}{\textsuperscript{58}} Department of Physics, University of Houston, Houston, TX 77204, USA\\
\hypertarget{aff:59}{\textsuperscript{59}} Department of Physics and Astronomy, Laurentian University, Sudbury, ON P3E 2C6, Canada\\
\hypertarget{aff:60}{\textsuperscript{60}} SNOLAB, Lively, ON P3Y 1N2, Canada\\
\hypertarget{aff:61}{\textsuperscript{61}} Department of Physics, Royal Holloway University of London, Egham, TW20 0EX, UK\\
\hypertarget{aff:62}{\textsuperscript{62}} Universit\`a degli Studi dell'Aquila, L'Aquila, 67100, Italy\\
\hypertarget{aff:63}{\textsuperscript{63}} Pharmacy Department, Universit\`a degli Studi ``Federico II'' di Napoli, Napoli, 80131, Italy\\
\hypertarget{aff:64}{\textsuperscript{64}} University of Warwick, Department of Physics, Coventry, CV47AL, UK\\
\hypertarget{aff:65}{\textsuperscript{65}} Amherst Center for Fundamental Interactions and Physics Department, University of Massachusetts, Amherst, MA 01003, USA\\
\hypertarget{aff:66}{\textsuperscript{66}} Department of Electrical and Electronic Engineering, Universit\`a degli Studi di Cagliari, Cagliari, 09123, Italy\\
\hypertarget{aff:67}{\textsuperscript{67}} Istituto Nazionale di Fisica Nucleare, Roma, 00186, Italia\\
\hypertarget{aff:68}{\textsuperscript{68}} School of Physics and Astronomy, University of Birmingham, Edgbaston, Birmingham, B15 2TT, UK\\
\hypertarget{aff:69}{\textsuperscript{69}} Physics Department, Lancaster University, Lancaster, LA1 4YB, UK\\
\hypertarget{aff:70}{\textsuperscript{70}} Center for Experimental Nuclear Physics and Astrophysics, and Department of Physics, University of Washington, Seattle, WA 98195, USAinstit\\
\hypertarget{aff:71}{\textsuperscript{71}} Chemical, Materials, and Industrial Production Engineering Department, Universit\`a degli Studi ``Federico II'' di Napoli, Napoli, 80126, Italy\\
\hypertarget{aff:72}{\textsuperscript{72}} Institute of High Energy Physics, Beijing, 100049, China\\
\hypertarget{aff:73}{\textsuperscript{73}} University of Chinese Academy of Sciences, Beijing, 100049, China\\
\hypertarget{aff:74}{\textsuperscript{74}} Department of Physics and Engineering, Fort Lewis College, Durango, CO 81301, USA\\
\hypertarget{aff:75}{\textsuperscript{75}} Institute of Applied Radiation Chemistry, Lodz University of Technology, Lodz, 93-590, Poland\\

\noindent
\thanks{$^\dagger$ Deceased}